\documentclass[12pt,prb,aps]{revtex4-1}
\usepackage{fullpage}
\usepackage{amssymb,amsmath}
\allowdisplaybreaks[1]
\usepackage{epsf}

\newcommand {\ltapp} {\stackrel {_{\normalsize<}}{_{\normalsize \sim}}}
\newcommand {\gtapp} {\stackrel {_{\normalsize>}}{_{\normalsize \sim}}}

\begin{document}
\title{\bf An Improved Neoclassical Drift-Magnetohydrodynamical Fluid Model of Helical Magnetic Island Equilibria in Tokamak Plasmas\\[0.5ex]
~\\[0.5ex]
{\rm Richard Fitzpatrick}\\[0.5ex]
{\it Institute for Fusion Studies}\\
{\it Department of Physics}\\
{\it University of Texas at Austin}\\~}
\begin{abstract}
The effect of the perturbed ion polarization current on the stability of 
neoclassical tearing modes is calculated using an improved,   neoclassical, four-field, drift-MHD model. The calculation involves the self-consistent
determination of the pressure and scalar electric potential profiles in the vicinity of the associated magnetic island chain, which allows  the  chain's propagation velocity to
be fixed. Two regimes are considered. First, a regime 
in which  neoclassical ion poloidal flow damping is not strong enough to enhance the magnitude of the polarization
current (relative to that found in slab geometry). Second, a regime  in which neoclassical
ion poloidal flow damping is strong enough to significantly enhance the magnitude of the polarization current. 
In both regimes, two types of solution are considered. First,   a freely rotating  solution (i.e., an island chain that
is not interacting with a static, resonant, magnetic perturbation). 
Second,  a locked solution (i.e., an island chain that has been brought to rest in the laboratory frame via interaction
with  a static, resonant, magnetic perturbation). In  all cases, the polarization current is found to be either always stabilizing, or stabilizing provided that
$\eta_i\equiv d\ln T_i/d\ln n_e$ does not exceed some threshold value. In certain ranges of $\eta_i$, the polarization current is found to have have a stabilizing effect
on a freely rotating island, but a destabilizing effect on a corresponding locked island.
\end{abstract}
\maketitle

\section{Introduction}
A tokamak is a device that is designed to trap a
thermonuclear plasma on a set of toroidally-nested magnetic flux-surfaces.\cite{wesson} Heat and particles are able to  flow  around the flux-surfaces
relatively rapidly due to the free streaming of charged particles
along magnetic field-lines. On the other hand, heat and
particles are only able to diffuse  across  the flux-surfaces relatively slowly, assuming that the magnetic field-strength
is large enough to render the particle gyroradii  much smaller
than the device's minor radius.\cite{boz}

Tokamak plasmas are subject to a number
of macroscopic  instabilities that limit their effectiveness.\cite{freidberg}
Such instabilities can be divided into two broad classes. So-called
 ideal instabilities are non-reconnecting modes that disrupt
the plasma in a matter of micro-seconds. However, such instabilities
can easily be avoided by limiting the plasma pressure and the net toroidal current.\cite{wes}  Tearing modes, on the other hand,
are relatively slowly growing instabilities that are  more difficult
to avoid.\cite{wes,fkr} These instabilities tend to saturate at relatively low levels,\cite{ruth,thy,esc,has} in the
process reconnecting magnetic flux-surfaces to form  helical
structures known as  magnetic island chains. Magnetic island chains 
are radially localized structures centered on so-called rational flux-surfaces, which
satisfy ${\bf k}\cdot{\bf B} = 0$, where ${\bf k}$ is the wave-number
of the instability, and ${\bf B}$ the equilibrium magnetic field. 
Island chains degrade plasma confinement because they enable heat and particles
to flow very rapidly along field-lines from their inner to their outer radii, implying an almost complete loss of confinement in the region lying between
these radii.\cite{chang}

As is well known, tearing mode dynamics in high-temperature tokamak plasmas is poorly described by the standard, single-fluid, resistive-magnetohydrodynamical (MHD) model.\cite{meiss}
 Indeed, in order to obtain realistic    predictions, at an absolute minimum, the resistive-MHD model must be replaced by a two-fluid, drift-MHD model. 
Broadly speaking,  the drift-MHD model predicts the existence of two separate branches of nonlinear tearing mode solutions.\cite{ott,mil,hyp,hyp1}
{\em Ion-branch}\/ solutions are characterized by  a flattened pressure profile within the island chain's magnetic
separatrix, a relatively large radial island width (compared to the poloidal ion gyroradius),  a propagation velocity similar to that of the
unperturbed local perpendicular ion fluid velocity, and no emission of drift-waves. On the other hand, {\em electron-branch}\/ solutions are characterized by a non-flattened pressure profile within the
magnetic separatrix, a relatively small radial width,  a propagation velocity close to that of the unperturbed local perpendicular electron fluid velocity, and the emission
of drift-waves. 
Numerical simulations suggest that the ion solution branch ceases to exist below a critical island width, whereas the 
electron solution branch ceases to exist above a second, somewhat larger, critical width.\cite{ott,mil} The disappearance of one branch of solutions is
associated with a bifurcation to the other branch.\cite{ott,mil}

This paper is concerned with the ion branch of nonlinear tearing mode solutions. The flattening of the pressure 
profile in the region lying within the island separatrix of such solutions gives rise to the disappearance of the neoclassical bootstrap current\,\cite{bic} there,
which has a strong destabilizing effect on the mode.\cite{car} Indeed, this effect is so marked that, unless countered, it would give rise to the formation of
magnetic island chains on every rational surface within the plasma, causing the complete destruction of magnetic flux-surfaces.\cite{heg}
In reality, this is not found to be the case. Instead, so-called {\em neoclassical tearing modes}\/ (i.e., tearing modes driven unstable by
the perturbed bootstrap current) are only observed to form on a few low mode-number rational surfaces within the plasma.\cite{chang1} This implies the existence of
a stabilizing mechanism that counters the destabilizing effect of the perturbed bootstrap current. Two possible mechanisms have been identified in the literature. 
First, the finite parallel transport in tokamak plasmas, combined with enhanced perpendicular transport due to plasma turbulence, may not allow the flattening of the
pressure profile within the magnetic separatrix.\cite{hel} However, this mechanism is only effective for relatively thin islands, and is  not
relevant to the ion solution branch.  The second stabilization mechanism, which appears to be the only feasible mechanism  for the ion branch, is associated with the perturbed ion polarization current.\cite{smol}

Calculating the effect of the perturbed ion polarization current on nonlinear tearing mode stability in a two-fluid plasma turns out to be a rather difficult task, for
a number of reasons. The first difficulty is that the sign of the polarization term in the island width evolution equation depends crucially on the island propagation velocity. Generally speaking, the
polarization term has one sign if the propagation velocity lies between the unperturbed local perpendicular guiding-center fluid velocity and the  unperturbed local 
perpendicular ion fluid velocity, and the opposite sign otherwise.\cite{smol,wil} Thus, a meaningful calculation of the polarization term must also be coupled
with a calculation of the island propagation velocity. The latter calculation involves a self-consistent determination of the pressure and scalar electric 
potential profiles in the vicinity of the island chain.\cite{f1,f2} The second difficulty is that the dominant contribution to the
polarization term originates from a boundary layer on the island chain's magnetic separatrix.\cite{flw} This contribution is such that the polarization term is
stabilizing when the island propagation velocity lies between the unperturbed local perpendicular guiding-center fluid velocity and the  unperturbed local 
perpendicular ion fluid velocity, and destabilizing otherwise.\cite{f1} If the contribution of the boundary layer is omitted then the sign of the
polarization term is reversed (so that the term is destabilizing  when the island propagation velocity lies between the unperturbed local perpendicular guiding-center fluid velocity and the unperturbed local 
perpendicular ion fluid velocity, and stabilizing otherwise).\cite{smol,wil} Unfortunately,  the contribution of the separatrix boundary layer to the polarization term is
a very sensitive function of the thickness of the layer.\cite{james1,james2} 
The final difficulty is that the magnitude of the polarization term is profoundly
affected by neoclassical ion poloidal flow damping.\cite{stix} Indeed, if the damping is sufficiently large then it gives rise to a coupling of the
perpendicular and parallel ion flows that acts to significantly enhance the magnitude of the polarization term.\cite{rob,re}

 Incidentally, because ion-branch magnetic islands are much wider than the poloidal ion gyroradius (and, hence, the ion banana width), it is reasonable to
assume that the response of both trapped and passing ions to the perturbed electric and magnetic fields in the vicinity of the island chain can be
adequately captured by a fluid model. Of course, such an assumption would not be not reasonable for island chains whose widths are comparable to, or
less than, the ion poloidal gyroradius.\cite{poli,berg,cai}

The aim of this paper is to present a two-fluid calculation of the ion polarization term appearing in the island width evolution equation  of a neoclassical
tearing mode in a high-temperature tokamak plasma. The calculation is performed using a neoclassical,  four-field, drift-MHD model. The
model itself was developed, and gradually improved, in Refs.~\onlinecite{rf1,rf2,rf3}. The core of the model is a single-helicity version of the
well-known four-field model of Hazeltine, Kotschenreuther, and Morrison.\cite{hkm} The core model is augmented by phenomenological terms
representing anomalous cross-field particle and momentum transport due to small-scale plasma turbulence. Finally, the model includes approximate (i.e.,
flux-surface averaged) expressions for the divergence of the neoclassical ion and electron stress tensors. These expressions allow us
to incorporate the bootstrap current, as well as neoclassical ion poloidal and perpendicular flow damping, into the model. Note that perpendicular
flow damping, which is due to nonambipolar transport associated with the breaking of toroidal symmetry by the tearing perturbation (and, possibly, by
external magnetic perturbations),\cite{sh} is often referred to in the literature as ``toroidal'' flow damping. This name is somewhat misleading, because the
damping actually acts on the perpendicular component of the ion fluid velocity. 

This paper is organized as follows. The neoclassical, four-field,  drift-MHD model that forms the basic of our analysis is introduced in Sect.~\ref{s2}. In Sect.~\ref{s3},
we calculate the ion polarization term for the case in which the neoclassical ion poloidal flow damping is not large enough to enhance the term's magnitude. In Sect.~\ref{s4}, we calculate the polarization term in the opposite case in which the flow damping is large enough
to significantly enhance the term's magnitude. The paper is summarized in Sect.~\ref{s5}. 

The general form of the calculations outlined in Sects.~\ref{s3} and \ref{s4} is 
similar to those described in Ref.~\onlinecite{rf2}. However, many of the details of the calculations are significantly modified by the improvements in the
expressions for the divergences of the neoclassical stress tensors introduced in Ref.~\onlinecite{rf3}. These improvements are as follows.
First, we have taken into account the fact that the neoclassical velocities towards which the divergences of the neoclassical stress tensors 
relax the electron and ion velocities are proportional to local electron and ion temperature gradients, respectively, and are, therefore, affected by
the modifications to these gradients induced by the presence of the island chain. Second, we have taken into account the fact that that the divergence of the
neoclassical ion perpendicular stress tensor generates a force that is primarily directed perpendicular to magnetic field-lines (within a given
flux-surface), rather than in the toroidal direction. In addition, we have  incorporated magnetic field-line curvature, the bootstrap current, and
independent equilibrium number density, electron temperature, and ion temperature gradients into the model. 

\section{Preliminary Analysis}\label{s2}
\subsection{Fundamental Definitions}
Consider a large aspect-ratio, low-$\beta$, circular cross-section,
 tokamak plasma equilibrium of major radius $R_0$, and toroidal magnetic field-strength $B_0$.
Let us adopt a right-handed, quasi-cylindrical, toroidal coordinate system ($r$, $\theta$, $\varphi$), whose symmetry axis ($r=0$) coincides with the
magnetic axis.  The coordinate $r$ also serves as a label
for the unperturbed (by the island chain) magnetic flux-surfaces. Let the equilibrium toroidal magnetic field and toroidal plasma current both run in the $+\varphi$ direction. 

 Suppose that a helical magnetic island chain,
with $m_\theta$ poloidal periods, and $n_\varphi$ toroidal periods, is embedded in
 the aforementioned plasma. The island chain is assumed to be  radially localized in the vicinity of its
associated 
rational surface, minor radius $r_s$,  which is defined as the unperturbed magnetic flux-surface at which $q(r_s)=m_\theta/n_\varphi$. Here, $q(r)$ is the   
safety-factor profile (which is assumed to be a monotonically increasing function of $r$). 
Let the full radial width of the island chain's magnetic separatrix be $4 \,w$.
In the following, it is assumed that
$r_s/R_0\ll 1$ and $w/r_s\ll 1$. 

The plasma is conveniently divided into an inner region, that comprises the plasma in the immediate vicinity of the
rational surface (and includes the island chain), and an outer region that comprises the remainder of the plasma. 
As is well known, in a high-temperature tokamak plasma, linear, ideal, MHD analysis invariably suffices to calculate the mode structure in the outer region, whereas nonlinear, nonideal, drift-MHD analysis
is generally required in the inner region. Let us assume that the linear, ideal, MHD solution has been found in the outer region. 
In the absence of an external perturbation, such a solution is characterized by a single  real parameter, ${\mit\Delta}'$,  (with units of inverse length)
known as the tearing stability index.\cite{fkr} 
The tearing stability index measures the free energy available in the outer region to
cause a spontaneous change in the island chain's radial width. This free energy acts to increase the width if ${\mit\Delta}'>0$,
and vice versa. It remains to obtain a nonlinear, nonideal, drift-MHD solution in the inner region, and then to
asymptotically match this solution to the aforementioned linear, ideal, MHD solution at the boundary between the inner and outer regions. 

All fields in the inner region are assumed to depend only on the  normalized radial
coordinate $X=(r-r_s)/w$, and the helical angle $\zeta=m_\theta\,\theta-n_\varphi\,\varphi-\phi_p(t)$. In particular, the
electron number density, electron temperature, and ion temperature profiles  in the inner region 
 take the forms $n(X,\zeta)=n_0\,(1+\delta n/n_0)$, $T_e(X,\zeta)=T_{e\,0}\,(1+\eta_e\,\delta n/n_0)$, and
$T_i(X,\zeta)=T_{i\,0}\,(1+\eta_i\,\delta n/n_0)$, respectively, Here, $n_0$, $T_{e\,0}$, $T_{i\,0}$, $\eta_e$, and $\eta_i$ are uniform
constants. Moreover, $\delta n(X,\zeta)/n_0\rightarrow - (w/L_n)\,X$ as $|X|\rightarrow \infty$, where $L_n>0$ is the density scale-length at the rational
surface. Note that we are assuming, for the sake of simplicity, that $\delta T_e/T_{e\,0}= \eta_e\,\delta n/n_0$, and $\delta T_i/T_{i\,0}= \eta_i\,\delta n/n_0$,
where $\delta T_e = T_e-T_{e\,0}$, et cetera. It follows that the flattening of the electron density profile within the island separatrix also implies the
flattening of the electron and ion temperature profiles. This approach is suitable for relatively wide, ion-branch magnetic island chains, where we expect complete
flattening of the pressure profile within the island separatrix, but would not be suitable for relatively narrow, electron-branch island chains, where we expect the electron
temperature profile to be flattened, but not the electron density and ion temperature profiles.\cite{hyp,hyp1} 

It is convenient to define the poloidal wavenumber, $k_\theta=m_\theta/r_s$, the resonant safety-factor, $q_s=m_\theta/n_\varphi$, 
the inverse aspect-ratio, $\epsilon_s=r_s/R_0$, the ion diamagnetic speed,
\begin{equation}
V_{\ast\,i}= \frac{T_{i\,0}\,(1+\eta_i)}{e\,B_0\,L_n},
\end{equation}
the electron diamagnetic speed, $V_{\ast\,e}= \tau\,V_{\ast\,i}$, where
\begin{equation}
\tau = \left(\frac{T_{e\,0}}{T_{i\,0}}\right)\left(\frac{1+\eta_e}{1+\eta_i}\right),
\end{equation}
the poloidal ion gyroradius, 
\begin{equation}
\rho_{\theta\,i} =\left(\frac{q_s}{\epsilon_s}\right)\left[\frac{T_{i\,0}\,(1+\eta_i)}{m_i}\right]^{1/2}\left(\frac{m_i}{e\,B_0}\right),
\end{equation}
and the ion beta, 
\begin{equation}
\beta_i =\frac{\mu_0\,n_0\,T_{i\,0}\,(1+\eta_i)}{B_0^{\,2}}.
\end{equation}
All of these quantities are evaluated at the rational surface. Here, $e$ is the magnitude of the electron charge, and $m_i$ the ion mass. 
Incidentally, the ions are assumed to be singly charged. 

\subsection{Fundamental Fields}
The  fundamental dimensionless fields in our  neoclassical, four-field, drift-MHD model are\,\cite{rf2,rf3}
\begin{align}
\psi(X,\zeta) &= \frac{q_s}{\epsilon_s}\,\frac{L_q}{w}\,\frac{A_\parallel}{B_0\,w},\\[0.5ex]
N(X,\zeta) &=\frac{L_n}{w}\,\frac{\delta n}{n_0},\\[0.5ex]
\phi(X,\zeta)&=-\frac{{\mit\Phi}}{w\,B_0\,V_{\ast\,i}}+v_p\,X,\\[0.5ex]
V(X,\zeta)&= \frac{\epsilon_s}{q_s}\,\frac{V_{\parallel\,i}}{V_{\ast\,i}}+v_p,
\end{align}
where 
\begin{align}
L_q &= 1\left/\left(\frac{d\ln q}{dr}\right)_{r=r_s}\right.,\\[0.5ex]
v_p &= \frac{1}{k_\theta\,V_{\ast\,i}}\,\frac{d\phi_p}{dt}.
\end{align}
Here, $A_\parallel$ is the component of the magnetic vector potential parallel to the equilibrium magnetic field (at the rational surface), 
$L_q>0$ the safety-factor scale-length at the rational surface, ${\mit\Phi}$  the electric scalar potential,  $v_p$  the normalized island
phase-velocity (which is assumed to be constant in time), and $V_{\parallel\,i}$ the component of the ion fluid velocity parallel to the equilibrium magnetic field
(at the rational surface).  The four fundamental fields are the normalized helical magnetic
flux, the normalized perturbed electron number density, the normalized electric scalar potential, and the normalized
parallel ion velocity, respectively. The four fundamental fields are evaluated in a frame of reference that moves with velocity $-(q_s/\epsilon_s)\,v_p\,V_{\ast\,i}\,{\bf e}_\varphi
=k_\varphi^{\,-1}\,(d\phi_p/dt)\,{\bf e}_\varphi$
with respect to the laboratory frame, where ${\bf e}_\varphi$ is a unit vector pointing in the $\varphi$-direction, and $k_\varphi = -n_\varphi/R_0$
 the toroidal wavenumber.  

\subsection{Neoclassical Four-Field Drift-MHD Model}
In the inner region, our  neoclassical, four-field, drift-MHD model  takes the form\,\cite{furuya,rf1,rf2,rf3}
\begin{align}
0&= [\phi+\tau\,N,\psi]+\beta\,\eta\,J\nonumber\\[0.5ex]
&\phantom{=}+\alpha_n^{-1}\,\hat{\nu}_{\theta\,e}\left[\alpha_n^{-1}\,J +V-\partial_X(\phi+\tau\,v_{\theta\,e}\,N)
-v_{\theta\,i}-\tau\,v_{\theta\,e}\right],\label{e11}\\[0.5ex]
0&= [\phi,N]-\rho\,[\alpha_n\,V+J,\psi]-\alpha_c\,\rho\,[\phi+\tau\,N,X]+D\,\partial_X^{\,2}N,\label{e12}\\[0.5ex]
0&=[\phi,V]-\alpha_n\,(1+\tau)\,[N,\psi]+\mu\,\partial_X^{\,2}V -\hat{\nu}_{\theta\,i}\!\left[V-\partial_X(\phi-v_{\theta\,i}\,N)\right],\label{e13}\\[0.5ex]
0&= \epsilon\,\partial_X[\phi-N,\partial_X \phi]+[J,\psi] +\alpha_c\,(1+\tau)\,[N,X]+\epsilon\,\mu\,\partial_X^{\,4}(\phi-N)\nonumber\\[0.5ex]
&\phantom{=}+\hat{\nu}_{\theta\,i}\,\partial_X\!\left[V -\partial_X(\phi-v_{\theta\,i}\,N)\right]+\hat{\nu}_{\perp\,i}\,\partial_X\!\left[-\partial_X(\phi
-v\,N)\right],\label{e14}
\end{align}
where
\begin{align}
J &= \beta^{\,-1}\left(\partial^{\,2}_X\psi-1\right),\label{e15}\\[0.5ex]
[A,B] &\equiv \partial_X A\,\,\partial_\zeta B -\partial_\zeta A\,\,\partial_X B.
\end{align}
Furthermore, $\partial_X\equiv (\partial/\partial X)_\zeta$ and $\partial_\zeta\equiv (\partial/\partial\zeta)_X$. 
Here, Eq.~(\ref{e11}) is the parallel Ohm's law, Eq.~(\ref{e12}) the electron continuity equation, Eq.~(\ref{e13}) the parallel ion
equation of motion, and Eq.~(\ref{e14}) the parallel ion vorticity equation. The auxiliary field $J(X,\zeta)$ is the normalized perturbed
parallel current. 

The various dimensionless parameters appearing in Eqs.~(\ref{e11})--(\ref{e15}) have the following definitions:
\begin{align}
\epsilon &=\left(\frac{\epsilon_s}{q_s}\right)^2,\\[0.5ex]
\rho&= \left(\frac{\rho_{\theta\,i}}{w}\right)^2,\\[0.5ex]
\alpha_n &= \frac{L_n/L_q}{\rho},\\[0.5ex]
\alpha_c&=\frac{2\,(L_n/L_c)}{\rho},\\[0.5ex]
\beta &= \frac{\beta_i}{\epsilon\,\rho\,\alpha_n^{\,2}},
\end{align}
and
\begin{align}
\eta&=\frac{\eta_{\parallel}}{\mu_0\,k_\theta\,V_{\ast\,i}\,w^{\,2}},\\[0.5ex]
D &= \left[D_\perp +\beta_i\,(1+\tau)\,\frac{\eta_\perp}{\mu_0}\left(1-\frac{3}{2}\,\frac{\eta_e}{1+\eta_e}\,\frac{\tau}{1+\tau}\right)\right]
\frac{1}{k_\theta\,V_{\ast\,i}\,w^{\,2}},\\[0.5ex]
\mu &= \frac{\mu_{\perp\,i}}{n_0\,m_i\,k_\theta\,V_{\ast\,i}\,w^{\,2}},
\end{align}
and
\begin{align}
\hat{\nu}_{\theta\,i}&=\left(\frac{\epsilon_s}{q_s}\right)^{\,2}\left(\frac{\nu_{\theta\,i}}{k_\theta\,V_{\ast\,i}}\right),\\[0.5ex]
\hat{\nu}_{\perp\,i}&=\left(\frac{\epsilon_s}{q_s}\right)^{\,2}\left(\frac{\nu_{\perp\,i}}{k_\theta\,V_{\ast\,i}}\right),\\[0.5ex]
\hat{\nu}_{\theta\,e}&=\left(\frac{m_e}{m_i}\right)\left(\frac{\epsilon_s}{q_s}\right)^{\,2}\left(\frac{\nu_{\theta\,e}}{k_\theta\,V_{\ast\,i}}\right),
\end{align}
and 
\begin{align}
v_{\theta\,i} &= 1+\lambda_{\theta\,i}\left(\frac{\eta_i}{1+\eta_i}\right)= 1-1.172\left(\frac{\eta_i}{1+\eta_i}\right),\label{e28h}\\[0.5ex]
v_{\perp\,i} &= 1+\lambda_{\perp\,i}\left(\frac{\eta_i}{1+\eta_i}\right)=1-2.367\left(\frac{\eta_i}{1+\eta_i}\right),\label{e29h}\\[0.5ex]
v_{\theta\,e} &=1- \lambda_{\theta\,e}\left(\frac{\eta_e}{1+\eta_e}\right)=1-0.717\left(\frac{\eta_e}{1+\eta_e}\right),\label{e30h}
\end{align}
and, finally, 
\begin{equation}\label{e44}
v = v_{\perp\,i}-v_p.
\end{equation}

Here,  $m_e$ is the electron mass, and $L_c$  the mean radius of curvature of magnetic field-lines at the rational surface.\cite{furuya,rf2} The
mean curvature is assumed to be favorable (i.e., $L_c>0$).\cite{kot} Note that we are neglecting the geodesic curvature of
magnetic field-lines, because this effect cannot be dealt with within the context of a single-helicity calculation. 

The quantities $\eta_\parallel$ and $\eta_\perp$ are the parallel and perpendicular plasma resistivities, respectively, whereras $D_\perp$
is a phenomenological perpendicular particle diffusivity (due to small-scale plasma turbulence), and $\mu_{\perp\,i}$ a phenomenological
perpendicular ion viscosity (likewise, due to small-scale turbulence). All four of these quantities are evaluated at the rational surface, and are assumed to
be constant across the inner region. 

\subsection{Model Neoclassical Stress Tensors}
The divergence of our model neoclassical ion stress tensor [which is used in the derivation of Eqs.~(\ref{e11})--(\ref{e15})]  takes the form\,\cite{furuya,rf3}
\begin{equation}
\nabla\cdot \pi_i = m_i\,n_0\left[\nu_{\theta\,i}\,V_{\theta\,i}^{nc}\,{\bf e}_\theta+\nu_{\perp\,i}\,V_{\perp\,i}^{nc}\,{\bf e}_\perp\right],
\end{equation}
where ${\bf e}_\perp= {\bf e}_\theta-(\epsilon_s/q_s)\,{\bf e}_\varphi$. Here, ${\bf e}_\theta$ is a unit vector
pointing in the $\theta$-direction, whereas ${\bf e}_\perp$ is a unit vector directed perpendicular to the
equilibrium magnetic field (at the rational surface). Moreover, $\nu_{\theta\,i}$ and $\nu_{\perp\,i}$ are the neoclassical
ion poloidal and perpendicular damping rates, respectively. Finally,
\begin{align}
V_{\theta\,i}^{nc} &= {\bf e}_\theta\cdot\left[{\bf V}_i - (1-v_{\theta\,i})\,{\bf V}_{\ast\,i}\right],\\[0.5ex]
V_{\perp\,i}^{nc} &={\bf e}_\perp\cdot\left[{\bf V}_i -(1- v_{\perp\,i})\,{\bf V}_{\ast\,i}-v_p\left({\bf V}_{\ast\,i}-{\bf V}_{\ast\,i}^{\,(0)}\right)\right],
\end{align}
Here, ${\bf V}_i$ is the ion fluid velocity (in the laboratory frame),  ${\bf V}_{\ast\,i}\equiv (\partial_X N)\,V_{\ast\,i}\,{\bf e}_\perp$ the ion diamagnetic
velocity, and  ${\bf V}_{\ast\,i}^{(0)}\equiv-V_{\ast\,i}\,{\bf e}_\perp$ the unperturbed (by the island chain) ion diamagnetic
velocity.

Note that (in the absence of the island chain) the neoclassical ion stress tensor acts to relax the ion poloidal fluid velocity in the vicinity of the rational
surface 
to the neoclassical value
\begin{equation}\label{e35t}
V_{\theta\,i} = (v_{\theta\,i}-1)\,V_{\ast\,i} = \lambda_{\theta\,i}\left(\frac{\eta_i\,V_{\ast\,i}}{1+\eta_i}\right)=-1.172\left(\frac{\eta_i\,T_{i\,0}}{e\,B_0\,L_n}\right), 
\end{equation}
and the ion perpendicular fluid velocity to the neoclassical value
\begin{equation}\label{e36t}
V_{\perp\,i} = (v_{\perp\,i}-1)\,V_{\ast\,i} = \lambda_{\perp\,i}\left(\frac{\eta_i\,V_{\ast\,i}}{1+\eta_i}\right)=-2.367\left(\frac{\eta_i\,T_{i\,0}}{e\,B_0\,L_n}\right).
\end{equation}
Inside the island separatrix (where ${\bf V}_{\ast\,i}={\bf 0}$, due to the flattening of the pressure profile), the neoclassical ion stress tensor
acts to relax the ion poloidal fluid velocity to zero, so that the island chain is convected by a purely toroidal flow. 

Neglecting the effect of plasma impurities, and assuming that the ions lie in the banana collisionality regime,
standard neoclassical theory yields $\nu_{\theta\,i}\sim \epsilon_s^{\,1/2}\nu_i/\epsilon$ and $\lambda_{\theta\,i}= -1.172$, where $\nu_i$ is the ion collision frequency.\cite{kim} 
Futhermore, assuming that the ion perpendicular flow damping lies in the so-called
``$1/\nu$ regime", neoclassical theory gives $\nu_{\perp\,i}\sim \epsilon_s^{\,3/2}\,n_\varphi^{\,2}\,(T_{i\,0}/m_i)\,(w/R_0)^{\,2}/(\epsilon\,R_0^{\,2}\,\nu_i)$ and
$\lambda_{\perp\,i}= -2.367$.\cite{sh0,sh}

The divergence of our model neoclassical electron stress tensor takes the form 
\begin{equation}
\nabla\cdot \pi_e = m_e\,n_0\,\nu_{\theta\,e}\,V_{\theta\,e}^{nc}\,{\bf e}_\theta,
\end{equation}
where $\nu_{\theta\,e}$ is the neoclassical electron poloidal flow damping rate, and 
\begin{equation}
V_{\theta\,e}^{nc} = {\bf e}_\theta\cdot\left[{\bf V}_e - (1-v_{\theta\,e})\,{\bf V}_{\ast\,e}\right].
\end{equation}
Here, ${\bf V}_e$ is the electron fluid velocity, and ${\bf V}_{\ast\,e}\equiv -\,(\partial_X N)\,V_{\ast\,e}\,{\bf e}_\perp$ the electron diamagnetic
velocity. Incidentally, $\nabla\cdot \pi_e$ is neglected with respect to $\nabla\cdot \pi_i$ when both appear in the same equation. 

Note that (in the absence of the island chain) the neoclassical electron stress tensor acts to relax the electron poloidal fluid velocity in the vicinity of the rational
surface 
to the neoclassical value 
\begin{equation}
V_{\theta\,e} = (1-v_{\theta\,e})\,V_{\ast\,e} = \lambda_{\theta\,e}\left(\frac{\eta_e\,V_{\ast\,e}}{1+\eta_e}\right)=0.717\left(\frac{\eta_e\,T_{e\,0}}{e\,B_0\,L_n}\right).
\end{equation}
Inside the island separatrix (where ${\bf V}_{\ast\,e}={\bf 0}$, due to the flattening of the pressure profile), the neoclassical electron stress tensor
acts to relax the electron poloidal fluid velocity to zero, 

Assuming that the electrons lie in the  banana collisionality regime,
 standard neoclassical theory yields $\nu_{\theta\,e}\sim \epsilon_s^{\,1/2}\,\nu_e/\epsilon$ and  $\lambda_{\theta\,e}=+ 0.717$, where
 $\nu_e$ is the electron collision frequency.\cite{pet} 
 
 Roughly speaking, our expressions for the divergences of the  neoclassical ion and electron stress tensors are the flux-surface averages of the
 true divergences. This approximate treatment of the divergences is necessary within the context of a single-helicity calculation. 

\subsection{Boundary Conditions}
Equations~(\ref{e11})--(\ref{e15}) are subject to the boundary conditions\,\cite{rf2,rf3}
\begin{align}
\psi(X,\zeta)&\rightarrow \frac{1}{2}\,X^{\,2}+ \cos\zeta,\label{e39}\\[0.5ex]
N(X,\zeta)&\rightarrow -X,\label{e40}\\[0.5ex]
\phi(X,\zeta)&\rightarrow -v\,X,\label{e41}\\[0.5ex]
V(X,\zeta)& \rightarrow v_{\theta\,i}-v,\label{e42}\\[0.5ex]
J(X,\zeta)&\rightarrow 0,\label{e43}
\end{align}
as $|X|\rightarrow \infty$.
It follows that the fields $\psi(X,\zeta)$, $V(X,\zeta)$, and $J(X,\zeta)$ are even in $X$, whereas the fields
$N(X,\zeta)$ and $\phi(X,\zeta)$ are odd. Of course, all fields are periodic in $\zeta$ with period $2\pi$. 

\subsection{Island Phase-Velocity Parameter}
The dimensionless parameter $v$, defined in Eq.~(\ref{e44}),  has a simple physical
interpretation. If $v=-\tau$ then the island chain co-rotates with the unperturbed electron fluid at the rational
surface; if $v=0$ then the island chain co-rotates with the unperturbed guiding-center fluid; and, if
$v=+1$ then the island chain co-rotates with the unperturbed ion fluid. 

\subsection{Island Geometry}
To lowest order, we expect that\,\cite{rf1,rf2,rf3}
\begin{equation}\label{e45e}
\psi(X,\zeta)={\mit\Omega}(X,\zeta)\equiv \frac{1}{2}\,X^{\,2}+\cos\zeta
\end{equation}
in the inner region. In fact, this result, which is known as the constant-$\psi$ approximation,\cite{fkr} holds as
long as $\beta\ll 1$. 
The contours of ${\mit\Omega}(X,\zeta)$ map out the magnetic flux-surfaces of a
helical magnetic island chain whose O-points are located at $X=0$ and $\zeta=\pi$, and
whose X-points are located at $X=0$ and $\zeta=0$. The magnetic separatrix corresponds to ${\mit\Omega}=1$,
the region enclosed by the separatrix to $-1\leq {\mit\Omega}<1$, and the region outside the separatrix to ${\mit\Omega}>1$. 

\subsection{Flux-Surface Average Operator}
The flux-surface average operator, $\langle\cdots\rangle$, is defined as the annihilator of $[A,{\mit\Omega}]$. In
other words, $\langle [A,{\mit\Omega}]\rangle =0$, for any field $A(X,\zeta)$. 
It follows that 
\begin{equation}
\langle A(s,{\mit\Omega},\zeta)\rangle =\oint \frac{A(s,{\mit\Omega},\zeta)}{[2\,({\mit\Omega}-\cos\zeta)]^{\,1/2}}\,\frac{d\zeta}{2\pi}
\end{equation}
for $1\leq {\mit\Omega}$, and 
\begin{equation}
\langle A(s,{\mit\Omega},\zeta)\rangle =\int_{\zeta_0}^{2\pi-\zeta_0} \frac{A(s,{\mit\Omega},\zeta)+A(-s,{\mit\Omega},\zeta) }{2\,[2\,({\mit\Omega}-\cos\zeta)]^{\,1/2}}\,\frac{d\zeta}{2\pi}
\end{equation}
for $-1\leq{\mit\Omega}<1$. Here, $s={\rm sgn}(X)$ and $\zeta_0=\cos^{-1}({\mit\Omega})$, where  $0\leq \zeta_0\leq \pi$. 

It is helpful to define
\begin{equation}
\tilde{A} \equiv A -\frac{\langle A\rangle}{\langle 1\rangle}.
\end{equation}
It follows that $\langle \tilde{A}\rangle=0$, for any field $A(X,\zeta)$. It is also easily
demonstrated that $\langle [A,F({\mit\Omega})]\rangle =0$, for any function $F({\mit\Omega})$.  

\subsection{Asymptotic Matching}
Standard asymptotic matching between the inner and outer regions\,\cite{rf0,ruth,ruth1} yields the island
width evolution equation,
\begin{equation}\label{e47}
4\,I_1\,\tau_R\,\frac{d}{dt}\!\left(\frac{w}{r_s}\right)= {\mit\Delta}'\,r_s +2\,m_\theta\left(\frac{w_v}{w}\right)^2\cos\phi_p + J_c\,\beta\,\frac{r_s}{w},
\end{equation}
and the island phase evolution equation, 
\begin{equation}\label{e48}
\frac{d^{\,2}\phi_p}{dt^{\,2}} \propto -2\,m_\theta\left(\frac{w_v}{w}\right)^2\sin\phi_p + J_s\,\beta\,\frac{r_s}{w}.
\end{equation}
Here, $I_1=0.823$ (see Appendix), $\tau_R = \mu_0\,r_s^{\,2}/\eta_\parallel$, and
\begin{align}\label{e49}
J_c&=-2\int_{-\infty}^{\infty} J\,\cos\zeta\,dX\,\frac{d\zeta}{2\pi} = -4\int_{-1}^\infty \langle J\,\cos\zeta\rangle\,d{\mit\Omega},\\[0.5ex]
J_s&=-2\int_{-\infty}^{\infty} J\,\sin\zeta\,dX\,\frac{d\zeta}{2\pi} = -4\int_{-1}^\infty \langle X\,[J,{\mit\Omega}]\rangle\,d{\mit\Omega}.\label{e50}
\end{align}

Note that, for the sake of completeness, we have taken into account the possibility that the plasma is subject
to a static, external, magnetic perturbation possessing the same helicity as the island chain. Here, $4\,w_v$ is the
full radial width of the vacuum island chain (i.e., the island chain obtained by naively superimposing the vacuum
magnetic perturbation onto the unperturbed plasma equilibrium), and $\phi_p$  becomes the helical phase-shift between the true island
chain and the vacuum island chain. 

The first term on the right-hand side of Eq.~(\ref{e47}) governs the intrinsic stability of the island
chain. (The chain is intrinsically stable if ${\mit\Delta}'<0$, and vice versa.) The second term represents the destabilizing effect
of the external perturbation. The final term represents the destabilizing or stabilizing (depending on whether the
integral $J_c$ is positive or negative, respectively) effect of helical currents flowing in the inner region. 

The first term on the right-hand side of Eq.~(\ref{e48}) represents the electromagnetic locking torque
exerted on the plasma in the inner region by the external perturbation. The second term represents the drag
torque due to the combined effects of neoclassical ion poloidal flow damping, neoclassical ion toroidal flow damping, and perpendicular ion
viscosity. 

\subsection{Expansion Procedure}
Equations~(\ref{e11})--(\ref{e15}) are solved, subject to the boundary conditions (\ref{e39})--(\ref{e43}), via an
expansion in two small parameters, $\Delta$ and $\delta$, where $\Delta\lll\delta\ll 1$. 
The expansion procedure is as follows. First, the coordinates $X$ and $\zeta$ are assumed to be ${\cal O}(\Delta^0\,\delta^{\,0})$. Next,  some particular ordering scheme is adopted for the fifteen physics parameters $v_{\theta\,i}$,  $v_{\theta\,e}$, $v$, $\tau$, $\alpha_n$, $\alpha_c$, $\epsilon$,
$\rho$, $\beta$,  $\hat{\nu}_{\theta\,i}$, $\hat{\nu}_{\perp\,i}$, $\hat{\nu}_{\theta\,e}$, $\eta$, $D$, 
and $\mu$. 
The fields $\psi$, $N$, $\phi$, $V$, and $J$
are then expanded in the form
$\psi(X,\zeta) =\sum_{i,j=0,\infty} \psi_{i,j}(X,\zeta)$,
et cetera, where $\psi_{i,j}\sim {\cal O}(\Delta^i\,\delta^{\,j})$.
Finally, Eqs.~(\ref{e11})--(\ref{e15})  are solved order by order, subject to the boundary conditions (\ref{e39})--(\ref{e43}).

\section{Weak Neoclassical Ion Poloidal Flow Damping Regime}\label{s3}
\subsection{Ordering Scheme}
The ordering scheme adopted in the so-called weak neoclassical ion poloidal flow damping regime is:\,\cite{rf2,rf3}
\begin{tabbing}
\mbox{\hspace{1cm}}\=$\Delta^0\,\delta^{\,0}$\=:~~~\=$v_{\theta\,i}$, $v_{\theta\,e}$, $v$, $\tau$, $\alpha_n$,\\[0.5ex]
\>$\Delta^0\,\delta^{\,1}$\>: \>$\alpha_c$, $\epsilon$, $\rho$, $\beta$,\\[0.5ex]
\>$\Delta^1\,\delta^{\,0}$\>: \>$\hat{\nu}_{\theta\,i}$, $\hat{\nu}_{\perp\,i}$, $\eta$, $D$, $\mu$,\\[0.5ex]
\>$\Delta^1\,\delta^{\,2}$\>: \>$\hat{\nu}_{\theta\,e}$.
\end{tabbing}
This ordering scheme is suitable for a constant-$\psi$ (i.e., $\beta\ll 1$) magnetic island chain whose radial width is much larger than the ion
poloidal gyroradius (i.e., $\rho\ll 1$), and which is embedded in a large aspect-ratio (i.e., $\epsilon\ll 1$), high-temperature (i.e., $\eta$, $D$, $\mu\lll 1$)
tokamak plasma equilibrium. The defining feature of the weak neoclassical ion poloidal flow damping regime is that the ion poloidal flow damping rate is sufficiently small that the neoclassical ion stress tensor is not the
dominant term in the ion parallel equation of motion. 

\subsection{Order $\Delta^0\,\delta^{\,0}$}
To order $\Delta^0\,\delta^{\,0}$, Eqs.~(\ref{e11})--(\ref{e15}) yield
\begin{align}
0&= [\phi_{0,0}+\tau\,N_{0,0},\psi_{0,0}],\label{e53}\\[0.5ex]
0&= [\phi_{0,0},N_{0,0}],\label{e54}\\[0.5ex]
0&=[\phi_{0,0}.V_{0,0}]-\alpha_n\,(1+\tau)\,[N_{0,0},\psi_{0,0}],\label{e55}\\[0.5ex]
0&=[J_{0,0},\psi_{0,0}],\label{e56}\\[0.5ex]
\partial_X^{\,2}\psi_{0,0}&=1.\label{e57}
\end{align}

Equations~(\ref{e39}), (\ref{e45e}), and  (\ref{e57}) give 
\begin{equation}\label{epo}
\psi_{0,0} = {\mit\Omega}(X,\zeta).
\end{equation}
Equations~(\ref{e40}), (\ref{e41}), (\ref{e53}), and (\ref{e54}) imply that
\begin{align}\label{e157}
\phi_{0,0} &= s\,\phi_0({\mit\Omega}),\\[0.5ex]
N_{0,0} &= s\,N_0({\mit\Omega}).\label{e158}
\end{align}
Note that, by symmetry, $\phi_0=N_0=0$ inside the separatrix, which means that the electron number density and temperature
profiles are flattened inside the separatrix. Let
\begin{align}\label{e59}
M({\mit\Omega}) &= -\frac{d\phi_0}{d{\mit\Omega}},\\[0.5ex]
L({\mit\Omega}) &= -\frac{dN_0}{d{\mit\Omega}}.\label{e60}
\end{align}
Equations~(\ref{e40}) and (\ref{e41}) yield
\begin{align}\label{e61}
M({\mit\Omega}\rightarrow\infty) &= \frac{v}{\sqrt{2\,{\mit\Omega}}},\\[0.5ex]
L({\mit\Omega}\rightarrow\infty) &= \frac{1}{\sqrt{2\,{\mit\Omega}}}.\label{e62}
\end{align}
Again, by symmetry, $M=L=0$ inside the separatrix. 
Equations~(\ref{e55}), (\ref{e59}), and (\ref{e60}) give
\begin{equation}\label{e63c}
V_{0,0} = V_0({\mit\Omega}),
\end{equation}
and Eq.~(\ref{e42}) implies that 
\begin{equation}\label{e64}
V_0({\mit\Omega}\rightarrow\infty) =  v_{\theta\,i}-v.
\end{equation}
Finally, Eq.~(\ref{e56}) yields
\begin{equation}
J_{0,0} = 0.\label{e67}
\end{equation}

\subsection{Order $\Delta^0\,\delta^{\,1}$}
To order $\Delta^0\,\delta^{\,1}$, Eqs.~(\ref{e14}),  (\ref{e157}), (\ref{e158}), and (\ref{e67}) give
\begin{equation}
[J_{1,0},{\mit\Omega}] = -\epsilon\,\partial_X[\phi_0-N_0,\partial_X\phi_0] -\alpha_c\,(1+\tau)\,[N_0,|X|].
\end{equation}
It follows, with the aid of Eqs.~(\ref{e59}) and (\ref{e60}), that 
\begin{equation}
[J_{0,1},{\mit\Omega}] =\left[\frac{\epsilon}{2}\,d_{\mit\Omega}\{(M-L)\,M\}X^{\,2}-\alpha_c\,(1+\tau)\,L\,|X|,{\mit\Omega}\right],
\end{equation}
where $d_{\mit\Omega}\equiv d/d{\mit\Omega}$. Hence, 
\begin{equation}\label{e70x}
J_{0,1} =\frac{\epsilon}{2}\,d_{\mit\Omega}[(M-L)\,M]\,\widetilde{X^{\,2}}-\alpha_c\,(1+\tau)\,L\,\widetilde{|X|} + \bar{J}({\mit\Omega}),
\end{equation}
where $\bar{J}({\mit\Omega})$ is an arbitrary flux function. 
However, the lowest-order flux-surface average of Eq.~(\ref{e11}) implies that
\begin{equation}\label{e71x}
\bar{J}({\mit\Omega}) =-\alpha_n\,\epsilon\,\nu_{\theta\,e}\,\tau_e\left(V_0 + \frac{M+\tau\,v_{\theta\,e}\,L}{\langle 1\rangle}-v_{\theta\,i}
-\tau\,v_{\theta\,e}\right),
\end{equation}
where
\begin{equation}
\tau_e = \nu_e^{\,-1}= \frac{m_e}{n_0\,e^{\,2}\,\eta_{\parallel}}
\end{equation}
is the electron collision time. 

Finally, it is easily demonstrated that
\begin{equation}
X\,[J_{0,1},{\mit\Omega}] = \frac{\epsilon}{6}\,[X^{\,3},(M-L)\,M] +\frac{1}{2}\,\alpha_c\,(1+\tau)\,[s\,X^{\,2},N_0],
\end{equation}
which implies that
\begin{equation}
\langle X\,[J_{0,1},{\mit\Omega}]\rangle = 0.
\end{equation}
In other words, $J_{0,1}$ does not contribute to the torque integral, $J_s$ [see Eq.~(\ref{e50})]. Thus,
in order to calculate $J_s$,  and, hence, to determine the phase-velocity of a freely rotating island chain, we must expand to higher order. 

\subsection{Order $\Delta^1\,\delta^{\,0}$}
To order $\Delta^1\,\delta^{\,0}$, Eqs.~(\ref{e11})--(\ref{e15}), (\ref{epo})--(\ref{e158}),  (\ref{e63c}), and (\ref{e67}) yield
\begin{align}
0&= [\phi_{1,0}+\tau\,N_{1,0},{\mit\Omega}] + s\,[\phi_0+\tau\,N_0,\psi_{1,0}],\label{e75}\\[0.5ex]
0&= s\,[\phi_{1,0},N_0] +s\,[\phi_0,N_{1,0}]+s\,D\,\partial_X^{\,2} N_0,\label{e76}\\[0.5ex]
0&= [\phi_{1,0},V_0] + s\,[\phi_0,V_{1,0}]-\alpha_n\,(1+\tau)\,[N_{1,0},{\mit\Omega}]-s\,\alpha_n\,(1+\tau)\,[N_0,\psi_{1,0}]
\nonumber\\[0.5ex]
&\phantom{=}+\mu\,\partial_X^{\,2}V_0-\hat{\nu}_{\theta\,i}\left[V_0-s\,\partial_X(\phi_0-v_{\theta\,i}\,N_0)\right],\label{e77}\\[0.5ex]
0&=[J_{1,0},{\mit\Omega}]+\hat{\nu}_{\theta\,i}\,\partial_X\!\left[V_0-s\,\partial_X(\phi_0-v_{\theta\,i}\,N_0)\right]
+\hat{\nu}_{\perp\,i}\,\partial_X\!\left[-s\,\partial_X(\phi_0-v\,N_0)\right]\label{e78},\\[0.5ex]
\partial_X^{\,2}\psi_{1,0}&=0.\label{e79}
\end{align}

It follows from Eq.~(\ref{e79}) that
\begin{equation}
\psi_{1,0} =0,
\end{equation}
from Eq.~(\ref{e75}) that
\begin{equation}
\phi_{1,0}=-\tau\,N_{1,0},
\end{equation}
from Eqs.~(\ref{e59}), (\ref{e60}), and (\ref{e76}) that
\begin{equation}\label{e82}
[N_{1,0},{\mit\Omega}] = D\left(\frac{X^{\,2}\,d_{\mit\Omega}L + L}{M+\tau\,L}\right),
\end{equation}
from Eq.~(\ref{e77}) that
\begin{align}\label{e83}
\left[\left\{\tau\,d_{\mit\Omega}V_0+\alpha_n\,(1+\tau)\right\}\!N_{1,0}-s\,M\,V_{1,0},{\mit\Omega}\right]
&=\mu\,X\,\partial_{\mit\Omega}(X\,d_{\mit\Omega}V_0)\nonumber\\[0.5ex]
&\phantom{=}
-\hat{\nu}_{\theta\,i}\left[V_0+|X|\,(M-v_{\theta\,i}\,L)\right],
\end{align}
and from Eq.~(\ref{e78}) that 
\begin{equation}\label{e84}
[J_{1,0},{\mit\Omega}]=-\hat{\nu}_{\theta\,i}\,\partial_X\!\left[V_0+|X|\,(M-v_{\theta\,i}\,L)\right]-\hat{\nu}_{\perp\,i}\,\partial_X\!
\left[
|X|\,(M-v\,L)\right].
\end{equation}
Here, $\partial_{\mit\Omega}\equiv(\partial/\partial{\mit\Omega})_\zeta$. 

Given that $M=L=0$ within the island separatrix, the previous four equations suggest that $\phi_{1,0}=N_{1,0}=V_{1,0}=J_{1,0}=V_0=0$ in this region. 
In particular, the flux-surface average of Eq.~(\ref{e84}) implies that $d_{\mit\Omega}V_0=0$ within the separatrix. The flux-surface average of Eq.~(\ref{e83})
then reveals that $V_0=0$ in this region. 

The flux-surface average of Eq.~(\ref{e82}) yields
\begin{equation}\label{e85e}
L({\mit\Omega}) = \left\{
\begin{array}{lll} 1/\langle X^{\,2}\rangle&\mbox{\hspace{0.5cm}}&1\leq{\mit\Omega}\\[0.5ex]
0&&-1\leq{\mit\Omega}<1\end{array}
\right.
\end{equation}

The flux-surface average of Eq.~(\ref{e84}) gives 
\begin{equation}
\hat{\nu}_{\theta\,i}\,d_{\mit\Omega}V_0 =-d_{\mit\Omega}\!\left[\hat{\nu}_{\theta\,i}\,\langle X^{\,2}\rangle\,(M-v_{\theta\,i}\,L)
+\hat{\nu}_{\perp\,i}\,\langle X^{\,2}\rangle\,(M-v\,L)\right]
\end{equation}
outside the magnetic separatrix. 
It follows from Eqs.~(\ref{e61}), (\ref{e62}),  (\ref{e64}), and  (\ref{e85e}) that 
\begin{equation}\label{e87}
V_0({\mit\Omega}) =-\left(\frac{\hat{\nu}_{\theta\,i}+\hat{\nu}_{\perp\,i}}{\hat{\nu}_{\theta\,i}}\right)
\left(\langle X^{\,2}\rangle\,F+\bar{v}\right)
\end{equation}
outside the separatrix, where 
\begin{equation}\label{e88}
\bar{v}= \frac{\hat{\nu}_{\theta\,i}\,(1-v_{\theta\,i})+\hat{\nu}_{\perp\,i}\,(1-v)}{\hat{\nu}_{\theta\,i}+\hat{\nu}_{\perp\,i}},
\end{equation}
and
\begin{equation}\label{e89}
F({\mit\Omega}) \equiv M({\mit\Omega})-L({\mit\Omega}).
\end{equation}
Note that $F=0$ inside the magnetic separatrix. The viscous term in Eq.~(\ref{e83}) requires continuity of $V_0({\mit\Omega})$ across the
separatrix. Given that $V_0=0$ inside the separatrix, and $\langle  X^{\,2}\rangle=4/\pi$ on the separatrix (see Appendix), Eq.~(\ref{e87}) yields 
\begin{equation}\label{e90}
F(1) =-\frac{\pi}{4}\,\bar{v}.
\end{equation}
Finally, Eqs.~(\ref{e61}), (\ref{e62}), and (\ref{e89}) give
\begin{equation}\label{e91}
F({\mit\Omega}\rightarrow\infty) = \frac{v-1}{\sqrt{2\,{\mit\Omega}}}.
\end{equation}

The flux-surface average of Eq.~(\ref{e83})  yields
\begin{equation}
0=\mu\,d_{\mit\Omega}\!\left(\langle X^{\,2}\rangle\,d_{\mit\Omega}V_0\right)
-\hat{\nu}_{\theta\,i}\,V_0\,\langle 1\rangle -\hat{\nu}_{\theta\,i}\left(F+
\frac{1-v_{\theta\,i}}{\langle X^{\,2}\rangle}\right)
\end{equation}
outside the magnetic separatrix. It follows from Eq.~(\ref{e87}) that\,\cite{rf2,rf3}
\begin{align}\label{e93}
0&=\hat{\mu}\,d_{\mit\Omega}\!\left[\langle X^{\,2}\rangle\,d_{\mit\Omega}(\langle X^{\,2}\rangle F)\right]-\hat{\nu}_{\theta\,i}\left(\langle X^{\,2}\rangle\langle 1\rangle
-1\right)\left(F+\frac{1-v_{\theta\,i}}{\langle X^{\,2}\rangle}\right)\nonumber\\[0.5ex]
&\phantom{=} -\hat{\nu}_{\perp\,i}\left(\langle X^{\,2}\rangle F+1-v\right)\langle 1\rangle,
\end{align}
where
\begin{equation}
\hat{\mu} =\left(\frac{\hat{\nu}_{\theta\,i}+\hat{\nu}_{\perp\,i}}{\hat{\nu}_{\theta\,i}}\right)\mu.
\end{equation}

\subsection{Evaluation of $J_c$}
According to Eqs.~(\ref{e70x}), (\ref{e71x}), (\ref{e85e}), and (\ref{e87})--(\ref{e89}), 
\begin{align}
J_{0,1} &= \frac{\epsilon}{2}\,d_{\mit\Omega}\!\left[F\left(F+\frac{1}{\langle X^{\,2}\rangle}\right)\right]\widetilde{X^{\,2}}
-\alpha_c\,(1+\tau)\,\frac{\widetilde{|X|}}{\langle X^{\,2}\rangle}\nonumber\\[0.5ex]
&\phantom{=} +\alpha_n\,\epsilon\,\nu_{\theta\,e}\,\tau_e\left[\left(
\langle X^{\,2}\rangle-\frac{1}{\langle 1\rangle}\right)\!F+\frac{\hat{\nu}_{\perp\,i}}{\hat{\nu}_{\theta\,i}}\left(\langle X^{\,2}\rangle F+1-v\right)\right.\nonumber\\[0.5ex]
&\phantom{=}\left.
+(1+\tau\,v_{\theta\,e})\left(1-\frac{1}{\langle 1\rangle\langle X^{\,2}\rangle}\right)\right]
\end{align}
for ${\mit\Omega}\geq 1$, and
\begin{equation}
J_{0,1} = \alpha_n\,\epsilon\,\nu_{\theta\,e}\,\tau_e\,(v_{\theta\,i}+\tau\,v_{\theta\,e})
\end{equation}
for $-1\leq {\mit\Omega}< 1$. 
Thus, it follows from Eqs.~(\ref{e45e}) and (\ref{e49})  that
\begin{equation}
J_c = J_p + J_g+ J_b,
\end{equation}
where
\begin{equation}\label{e98}
J_p =\epsilon\int_{1-}^{\infty} d_{\mit\Omega}\!\left[F\left(F+\frac{1}{\langle X^{\,2}\rangle}\right)\right]\langle \widetilde{X^{\,2}}\,\widetilde{X^{\,2}}\rangle\,d{\mit\Omega}
\end{equation}
parameterizes the effect of the perturbed ion polarization current on island stability,  whereas 
\begin{equation}
J_g = -\alpha_c\,(1+\tau)\int_1^{\infty}2\,\frac{\langle\widetilde{|X|}\,\widetilde{X^{\,2}}\rangle}{\langle X^{\,2}\rangle}\,d{\mit\Omega}
\end{equation}
parameterizes the effect of magnetic field-line curvature on island stability, and, finally, 
\begin{align}\label{e100}
J_b &=-\alpha_n\,\epsilon\,\nu_{\theta\,e}\,\tau_e\int_1^\infty 2\left\{\left(
\langle X^{\,2}\rangle-\frac{1}{\langle 1\rangle}\right)F+\frac{\hat{\nu}_{\perp\,i}}{\hat{\nu}_{\theta\,i}}\left(\langle X^{\,2}\rangle F+1-v\right)\right.\nonumber\\[0.5ex]
&\phantom{=}\left.
+(1+\tau\,v_{\theta\,e})\left(1-\frac{1}{\langle 1\rangle\langle X^{\,2}\rangle}\right)-v_{\theta\,i}-\tau\,v_{\theta\,e}\right\}\left(2\,{\mit\Omega}\,\langle 1\rangle-\langle X^{\,2}\rangle\right)
d{\mit\Omega}
\end{align}
parameterizes the effect of the perturbed bootstrap current on island stability. In deriving the previous expressions, we have made use of the
following easily demonstrated results: $\langle \widetilde{A}\,\cos\zeta\rangle = -(1/2)\,\langle\widetilde{A}\,\widetilde{X^{\,2}}\rangle$ and 
$\int_{-1}^\infty \langle \cos\zeta\rangle\,d{\mit\Omega}=0$. 

\subsection{Evaluation of $J_s$}
Equations~(\ref{e84}), (\ref{e85e}), and  (\ref{e87})--(\ref{e89}) imply that
\begin{equation}
[J_{1,0},{\mit\Omega}] =-\partial_X G,
\end{equation}
where 
\begin{align}
G= -\hat{\nu}_{\theta\,i}\left(\langle X^{\,2}\rangle -|X|\right)\left(F + \frac{1-v_{\theta\,i}}{\langle X^{\,2}\rangle}\right) - \hat{\nu}_{\perp\,i}\left(\langle X^{\,2}\rangle -|X|\right)\left(F + \frac{1-v}{\langle X^{\,2}\rangle}\right)
\end{align}
for ${\mit\Omega}\geq 1$, and $G=0$ for $-1\leq {\mit\Omega} < 1$. Note that $G$ is continuous across the separatrix (${\mit\Omega}=1$). 
It follows that\,\cite{rf2,rf3}
\begin{equation}
\langle X\,[J_{1,0},{\mit\Omega}]\rangle =-d_{\mit\Omega}\langle X^{\,2}\,G\rangle+\langle G\rangle.
\end{equation}
Hence, Eq.~(\ref{e50}) yields
\begin{equation}
J_s = -4\,\int_1^\infty \langle G\rangle \,d{\mit\Omega},
\end{equation}
because $\langle X^{\,2}\,G\rangle_{{\mit\Omega}\rightarrow\infty} = 0$. 
Thus, we obtain
\begin{align}\label{e105}
J_s&= \hat{\nu}_{\theta\,i}\int_1^{\infty}4\left(\langle 1\rangle\langle X^{\,2}\rangle -1\right)\left(F + \frac{1-v_{\theta\,i}}{\langle X^{\,2}\rangle}\right)d{\mit\Omega}
\nonumber\\[0.5ex]&\phantom{=} 
+ \hat{\nu}_{\perp\,i}\int_1^\infty4\left(\langle 1\rangle\langle X^{\,2}\rangle -1\right)\left(F + \frac{1-v}{\langle X^{\,2}\rangle}\right)d{\mit\Omega}.
\end{align}

\subsection{Transformed Equations}\label{strans}
Let 
\begin{equation}
Y(k) = -2\,k\left.\left[F(k) +\frac{1-v_{\theta\,i}}{2\,k\,{\cal C}(k)}\right]\right/(v_{\theta\,i}-v),
\end{equation}
where 
$k=[(1+{\mit\Omega})/2]^{\,1/2}$. Note that $k=0$ corresponds to the island O-point, $k=1$ to the island separatrix, and $k\rightarrow\infty$ to ${\mit\Omega}\rightarrow \infty$. 
Here, ${\cal C}(k)$ is defined in the Appendix. It follows from Eqs.~(\ref{e88}), (\ref{e90}),
and (\ref{e91}) that 
\begin{align}\label{e107}
Y(1)& =\frac{\pi}{2}\left(\frac{\hat{\nu}_{\perp\,i}}{\hat{\nu}_{\theta\,i}+\hat{\nu}_{\perp\,i}}\right),\\[0.5ex]
Y(k\rightarrow\infty) &=1.\label{e108}
\end{align}
Furthermore, Eq.~(\ref{e93}) reduces to
\begin{equation}\label{e109}
0=\frac{\hat{\mu}}{4}\,d_k\!\left[{\cal C}\,d_k({\cal C}\,Y)\right]-\hat{\nu}_{\theta\,i}\,({\cal A}\,{\cal C}-1)\,Y-\hat{\nu}_{\perp\,i}\,({\cal C}\,Y-1)\,{\cal A},
\end{equation}
where $d_k\equiv d/dk$, and  ${\cal A}(k)$ is defined in the Appendix. 
Equations~(\ref{e98})--(\ref{e100}) yield 
\begin{align}
J_p&= \epsilon\int_{1-}^{\infty}\frac{d}{dk}\!\left[\left\{(v_{\theta\,i}-v)\,\frac{Y}{2\,k}
+\frac{1-v_{\theta\,i}}{2\,k\,{\cal C}}\right\}\left\{(v_{\theta\,i}-v)\,\frac{Y}{2\,k}
-\frac{v_{\theta\,i}}{2\,k\,{\cal C}}\right\}\right]8\left({\cal E}-\frac{{\cal C}^{\,2}}{{\cal A}}\right)k^{\,3}\,dk,\label{e110}\\[0.5ex]
J_g &=-\alpha_c\,(1+\tau)\int_1^\infty 16\left(\frac{{\cal D}}{B}-\frac{1}{{\cal A}}\right)k^{\,2}\,dk,\\[0.5ex]
J_b&= \alpha_n\,\epsilon\,\nu_{\theta\,e}\,\tau_e\left[1-v-\tau\,(1-v_{\theta\,e})\right]\int_1^\infty 16\left[\left\{1+
\frac{\hat{\nu}_{\perp\,i}}{\hat{\nu}_{\theta\,i}}-\frac{1}{{\cal A}\,{\cal C}}\right\}{\cal C}\,Y-\frac{\hat{\nu}_{\perp\,i}}{\hat{\nu}_{\theta\,i}}\right]
({\cal D}\,{\cal A}-{\cal C})\,k^{\,2}\,dk\nonumber\\[0.5ex]
&\phantom{=}+ \alpha_n\,\epsilon\,\nu_{\theta\,e}\,\tau_e\,(v_{\theta\,i}+\tau\,v_{\theta\,e})\int_1^\infty 16\left(\frac{{\cal D}}{{\cal C}}-\frac{1}{{\cal A}}\right)
k^{\,2}\,dk,
\end{align}
where ${\cal D}(k)$ and ${\cal E}(k)$ are defined in the Appendix. Finally, Eq.~(\ref{e105}) gives
\begin{equation}\label{e113}
J_s =-\hat{\nu}_{\theta\,i}\,(v_{\theta\,i}-v)\int_1^\infty 8\,({\cal A}\,{\cal C} -1)\,Y\,dk-\hat{\nu}_{\perp\,i}\,(v_{\theta\,i}-v)
\int_1^\infty 8\,({\cal A}\,{\cal C}-1)\left(Y-\frac{1}{{\cal C}}\right)dk.
\end{equation}
It remains to solve Eq.~(\ref{e109}), subject to the boundary conditions (\ref{e107}) and (\ref{e108}), and then to evaluate the integrals (\ref{e110})--(\ref{e113}). 
This task, which involves the elimination of an unphysical solution that varies as
$Y\sim \exp[+2\,(\hat{\nu}_{\perp\,i}/\hat{\mu})^{1/2}\,k]$ as $k\rightarrow\infty$, can be achieved analytically in five different parameter regimes that
are described in Sect.~\ref{soln}.\cite{rf3} 

\subsection{Separatrix Boundary Layer}\label{sq}
The flux-surface functions $M({\mit\Omega})$ and $L({\mit\Omega})$  are both zero inside, and non-zero
just outside, the magnetic separatrix. 
Retaining selected higher-order terms (containing radial derivatives) in Eqs.~(\ref{e76}) and (\ref{e78}), we find that
\begin{align}
(M+\tau\,L)\,[N_{1,0},{\mit\Omega}]-s\,\rho\,[J_{1,0},{\mit\Omega}] &\simeq D\left(X^{\,2}\,d_{\mit\Omega} L + L\right),\\[0.5ex]
s\,[J_{1,0},{\mit\Omega}] &\simeq -\epsilon\,\partial_X[\phi_0-N_0, \partial _X \phi_{1,0}]=\epsilon\,\tau\,(M-L)\,\partial_X^{\,2}[N_{1,0},{\mit\Omega}],
\end{align}
so that Eq.~(\ref{e82}) generalizes to give
\begin{equation}
\left\{M+\tau\,L-\tau\,(M-L)\,\epsilon\,\rho\,\partial_X^{\,2}\right\}[N_{1,0},{\mit\Omega}]\simeq  D\left(X^{\,2}\,d_{\mit\Omega} L + L\right),
\end{equation}
which suggests that the apparent discontinuities in the functions $M({\mit\Omega})$ and $L({\mit\Omega})$  are
resolved in a thin boundary layer, centered on the magnetic separatrix, of (unnormalized) width $(\epsilon\,\rho)^{1/2}\,w=\rho_i$, where $\rho_i=(\epsilon_s/q_s)\,\rho_{\theta\,i}$ is the ion gyroradius.\cite{bound}

\begin{figure}
\epsfysize=4in
\centerline{\epsffile{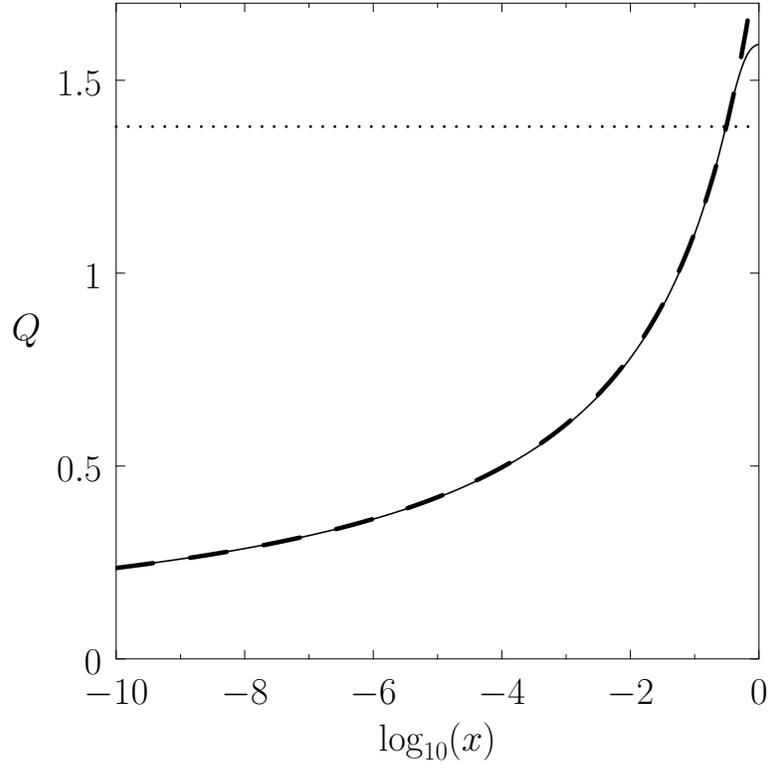}}
\caption{The solid curve shows the separatrix boundary layer response function, $Q(x)$. The
dashed curve shows the analytic approximation $Q(x)\simeq 6.2/
\ln(16/x)-3.0/\ln^{\,2}(16/x)$. The dotted line corresponds to $Q=1.38$.}\label{f0}
\end{figure}

Equation~(\ref{e110}) can be written
\begin{equation}\label{epolz}
J_p=\epsilon\int_{1-}^\infty d_k\left[F\,(F+L)\right]8\left({\cal E} - \frac{{\cal C}^{\,2}}{{\cal A}}\right)k^{\,3}\,dk.
\end{equation}
In accordance with Eqs.~(\ref{e85e}) and (\ref{e90}), let us suppose that, in the immediate vicinity of the separatrix, $L(k)$ and
$F(k)\equiv M(k)-L(k)$ take the following forms:
\begin{align}
L(k) &= \frac{f(k)}{2\,k\,{\cal C}(k)},\\[0.5ex]
F(k) &= -\frac{\bar{v}\,f(k)}{2\,k\,{\cal C}(k)},
\end{align}
where
\begin{equation}
f(k)= \frac{1}{2}\left(1+\tanh\left[(k-1)\,\frac{2\,w}{\rho_i}\right]\right).
\end{equation}
In effect, we have  resolved the discontinuities in the functions $L(k)$ and $F(k)$ across the separatrix in a  boundary layer of (unnormalized) thickness
$\rho_i$. In the limit that $\rho_i/w\ll 1$, the contribution of the boundary layer to the polarization integral (\ref{epolz}) can be written 
\begin{equation}
J_{p\,s} = \left[\frac{2\pi}{3}- Q\left(\frac{\rho_i}{w}\right)\right]\epsilon\,\bar{v}\,(\bar{v}-1),
\end{equation}
where
\begin{equation}
Q(x)=2\pi\int_0^\infty \frac{{\rm sech}^2(y)}{\ln(16/x)+\ln(1/y)}\,dy\simeq \frac{6.2}{\ln(16/x)} - \frac{3.0}{\ln^{\,2}(16/x)}.
\end{equation}
In deriving the previous equation, we have made use of the fact that the functions ${\cal C}(k)$ and ${\cal E}(k)$ are well behaved as $k\rightarrow 1$, whereas
the function ${\cal A}(k)$ has a logarithmic singularity.\cite{ab} The separatrix boundary layer response function, $Q(x)$, is plotted in Fig.~\ref{f0}.

According to the previous analysis, the polarization integral, (\ref{e110}),  takes the form
\begin{align}\label{e110x}
J_p&= 
\left[\frac{2\pi}{3}- Q\left(\frac{\rho_i}{w}\right)\right]\epsilon\,\bar{v}\,(\bar{v}-1)\\[0.5ex]
&\phantom{=}+\epsilon\int_{1+}^{\infty}\frac{d}{dk}\!\left[\left\{(v_{\theta\,i}-v)\,\frac{Y}{2\,k}
+\frac{1-v_{\theta\,i}}{2\,k\,{\cal C}}\right\}\left\{(v_{\theta\,i}-v)\,\frac{Y}{2\,k}
-\frac{v_{\theta\,i}}{2\,k\,{\cal C}}\right\}\right]8\left({\cal E}-\frac{{\cal C}^{\,2}}{{\cal A}}\right)k^{\,3}\,dk.\nonumber
\end{align}
 Of course, the first term on the right-hand side of the previous equation emanates from the separatrix boundary layer.\cite{flw} Note that the neglect of the
 finite thickness of the boundary layer leads to a significant overestimate of the contribution of the layer to the polarization integral.\cite{james1,james2}

\subsection{Island Solution Regimes}\label{soln}
\begin{figure}
\epsfysize=4in
\centerline{\epsffile{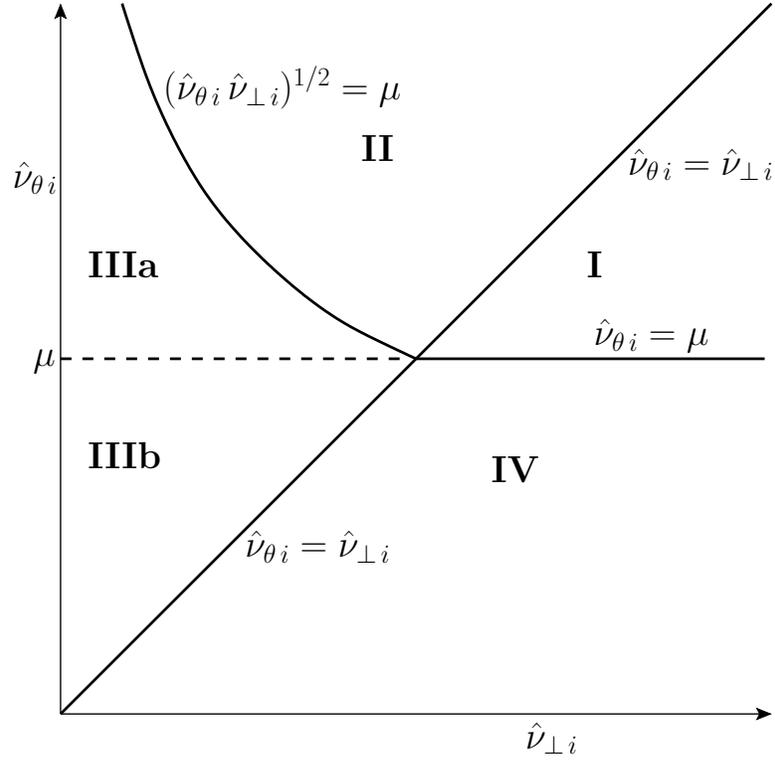}}
\caption{Extends of various weak neoclassical ion poloidal flow damping island solution regimes in the $\hat{\nu}_{\perp\i}$--$\hat{\nu}_{\theta\,i}$ plane. }\label{f1}
\end{figure}

The extents of the five analytic solution regimes, mentioned in Sect.~\ref{strans}, in the $\hat{\nu}_{\perp\,i}$--$\hat{\nu}_{\theta\,i}$ plane
are indicated in Fig.~\ref{f1}. 

Regime~I corresponds to $\hat{\nu}_{\perp\,i}\gg \hat{\nu}_{\theta\,i}$ and $\mu\ll \hat{\nu}_{\theta\,i}$. In this regime,
\begin{equation}
Y(k) \simeq \frac{1}{{\cal C}}\left[1-\frac{\hat{\nu}_{\theta\,i}}{\hat{\nu}_{\perp\,i}}\left(1-\frac{1}{{\cal A}\,{\cal C}}\right)\right].
\end{equation}
It follows that
\begin{align}
J_p &= -\left[I_2-Q\left(\frac{\rho_i}{w}\right)\right]\epsilon\,v\,(1-v),\\[0.5ex]
J_g &= -I_3\,\alpha_c\,(1+\tau),\\[0.5ex]
J_b &= I_3\,\alpha_n\,\epsilon\,\nu_{\theta\,e}\,\tau_e\,(v_{\theta\,i}+\tau\,v_{\theta\,e}),\\[0.5ex]
J_s &= -I_4\,\hat{\nu}_{\theta\,i}\,(v_{\theta\,i}-v),
\end{align}
where $I_2=1.38$, $I_3=1.58$, and $I_4=0.357$ are defined in the Appendix. 

Regime~II corresponds to $\hat{\nu}_{\theta\,i}\gg \hat{\nu}_{\perp\,i}$ and $\mu\ll (\hat{\nu}_{\theta\,i}\,\hat{\nu}_{\perp\,i})^{1/2}$. In this regime,
\begin{equation}
Y(k) \simeq \frac{\hat{\nu}_{\perp\,i}}{\hat{\nu}_{\theta\,i}}\,\frac{{\cal A}}{{\cal A}\,{\cal C}\,(1+\hat{\nu}_{\perp\,i}/\hat{\nu}_{\theta\,i})-1}.
\end{equation}
It follows that
\begin{align}
J_p &= -\left[I_2-Q\left(\frac{\rho_i}{w}\right)\right]\epsilon\,v_{\theta\,i}\,(1-v_{\theta\,i}),\\[0.5ex]
J_g &= -I_3\,\alpha_c\,(1+\tau),\\[0.5ex]
J_b &= I_3\,\alpha_n\,\epsilon\,\nu_{\theta\,e}\,\tau_e\,(v_{\theta\,i}+\tau\,v_{\theta\,e}),\\[0.5ex]
J_s &= -I_5\,\hat{\nu}_{\theta\,i}^{\,1/4}\,\hat{\nu}_{\perp\,i}^{\,3/4}\,(v_{\theta\,i}-v),
\end{align}
where $I_5= 3.74$ is defined in the Appendix. 

Regime~IIIa corresponds to $\hat{\nu}_{\theta\,i}\gg \mu$ and $\mu\gg (\hat{\nu}_{\theta\,i}\,\hat{\nu}_{\perp\,i})^{1/2}$.
In this regime,
\begin{equation}
Y(k)\simeq \frac{\hat{\nu}_{\perp\,i}}{\hat{\nu}_{\theta\,i}}\,\frac{{\cal A}}{{\cal A}\,{\cal C}\,(1+\hat{\nu}_{\perp\,i}/\hat{\nu}_{\theta\,i})-1}
\end{equation}
for $1<k\ll k_1$, and
\begin{equation}
Y(k)\simeq 1-\left(1+\frac{k_1}{k_2}\right){\rm e}^{-k/k_2}
\end{equation}
for $k\gtapp k_1$. Here, $k_1=(\hat{\nu}_{\theta\,i}/8\,\mu)^{\,1/2}$ and $k_2=(\mu/4\,\hat{\nu}_{\perp\,i})^{\,1/2}$. 
Regime~IIIb corresponds to $\hat{\nu}_{\theta\,i}\gg \hat{\nu}_{\perp\,i}$ and $\mu \gg \hat{\nu}_{\theta\,i}$.
In this regime,
\begin{equation}
Y(k)\simeq \frac{\hat{\nu}_{\perp\,i}}{\hat{\nu}_{\theta\,i}\,{\cal C}}
\end{equation}
for $1<k\ll k_3$, and
\begin{equation}
Y(k)\simeq 1-\left(1+\frac{k_3}{k_2}\right){\rm e}^{-k/k_2}
\end{equation}
for $k\gtapp k_3$, where $k_3=(\mu/4\,\hat{\nu}_{\theta\,i})^{\,1/2}$. In both regimes IIIa and IIIb, 
\begin{align}
J_p &= -\left[I_2-Q\left(\frac{\rho_i}{w}\right)\right]\epsilon\,v_{\theta\,i}\,(1-v_{\theta\,i}),\\[0.5ex]
J_g &= -I_3\,\alpha_c\,(1+\tau),\\[0.5ex]
J_b &= I_3\,\alpha_n\,\epsilon\,\nu_{\theta\,e}\,\tau_e\,(v_{\theta\,i}+\tau\,v_{\theta\,e}),\\[0.5ex]
J_s &= -4\,(\hat{\nu}_{\perp\,i}\,\mu)^{1/2}\,(v_{\theta\,i}-v).
\end{align}

Finally, Regime~IV corresponds to $\hat{\nu}_{\theta\,i}\ll \hat{\nu}_{\perp\,i}$ and $\mu\gg \hat{\nu}_{\theta\,i}$. In this regime,
\begin{equation}
Y(k) \simeq \frac{1}{{\cal C}}\left(1-\frac{\hat{\nu}_{\theta\,i}}{\hat{\nu}_{\perp\,i}}\,{\rm e}^{-k/k_3}\right).
\end{equation}
It follows that
\begin{align}
J_p &= -\left[I_2-Q\left(\frac{\rho_i}{w}\right)\right]\epsilon\,v\,(1-v),\\[0.5ex]
J_g &= -I_3\,\alpha_c\,(1+\tau),\\[0.5ex]
J_b &= I_3\,\alpha_n\,\epsilon\,\nu_{\theta\,e}\,\tau_e\,(v_{\theta\,i}+\tau\,v_{\theta\,e}),\\[0.5ex]
J_s &= -4\,(\hat{\nu}_{\theta\,i}\,\mu)^{\,1/2}\,(v_{\theta\,i}-v).
\end{align}

\subsection{Freely Rotating Magnetic Islands}\label{sfree}
Consider a freely rotating magnetic island chain: that is, a chain which is not interacting with a static, resonant, external, magnetic perturbation.
This implies  that $w_v=0$ in Eq.~(\ref{e48}). Hence, because we have already assumed that $d^{\,2}\phi_p/dt^{\,2}=0$ (i.e., the
island is rotating steadily), we conclude that $J_s=0$. In other words, there is zero local drag torque acting on a freely rotating island chain. 

According to the analysis in Sect.~\ref{soln}, a freely rotating island  chain is characterized by
\begin{equation}\label{e137}
v= v_{\theta\,i} = 1+\lambda_{\theta\,i}\left(\frac{\eta_i}{1+\eta_i}\right) = \frac{1-0.172\,\eta_i}{1+\eta_i},
\end{equation}
where use has been made of Eq.~(\ref{e28h}).
We conclude that the phase-velocity of a freely rotating chain is solely determined by the  neoclassical
ion poloidal velocity [which is parameterized by $v_{\theta\,i}$---see Eq.~(\ref{e35t})]. Moreover, the phase-velocity lies between the unperturbed local perpendicular
guiding-center fluid velocity and the unperturbed local perpendicular ion fluid velocity (i.e., $0<v<1$, as is seen experimentally\,\cite{lhaye})
provided that $0<\eta_i<5.81$. On the other hand, if $\eta_i>5.81$ then the chain rotates in the local electron diamagnetic direction (i.e., $v<0$).

The analysis in Sect.~\ref{soln} implies that
\begin{equation}\label{e138}
J_p = -\left[1.38-Q\left(\frac{\rho_i}{w}\right)\right]\epsilon\,v_{\theta\,i}\,(1-v_{\theta\,i})=-\left[1.38-Q\left(\frac{\rho_i}{w}\right)\right]\epsilon\,v\,(1-v),
\end{equation}
where use has been made of Eq.~(\ref{e137}). Now, it is clear from Fig.~\ref{f0} that $1.38-Q(\rho_i/w)>0$ unless $\rho_i/w\gtapp 0.3$. However, 
$\rho_i/w\gtapp 0.3$ then
is not consistent with an ion-branch magnetic island chain characterized by $w\gg \rho_{\theta\,i}$. 
We conclude that  the perturbed ion polarization current has a stabilizing effect on the island chain (i.e., $J_p<0$) provided that the chain's
phase-velocity lies between the unperturbed local perpendicular
guiding-center fluid velocity and the unperturbed local perpendicular ion fluid velocity (i.e., $0<v<1$). As we have just seen, this is
the case as long as $0<\eta_i<5.81$. On the other hand, if $\eta_i>5.81$ then $v<0$, and the polarization term becomes destabilizing. 

The analysis in Sect.~\ref{soln} yields
\begin{equation}
J_g= -1.58\,\alpha_c\,(1+\tau).
\end{equation}
In other words, magnetic field-line curvature has a stabilizing effect on the island chain (i.e., $J_g<0$).

Finally, the analysis in Sect.~\ref{soln} implies that 
\begin{equation}
J_b = 3.85\,\epsilon_s^{\,1/2}\,\alpha_n\left[\frac{(1-0.172\,\eta_i)\,T_{i\,0}+(1+0.283\,\eta_e)\,T_{e\,0}}{(1+\eta_i)\,T_{i\,0}}\right],
\end{equation}
where we have made use of 
 Eqs.~(\ref{e28h}) and (\ref{e30h}), as well as the standard neoclassical result $\epsilon\,\nu_{\theta\,e}\,\tau_e = 1.67\,f_t$, where $f_t\simeq 1.46\,\epsilon_s^{\,1/2}$
is the faction of trapped particles at the rational surface.\cite{kim,pet} It follows that the perturbed bootstrap current is
destabilizing (i.e., $J_b>0$) provided that
\begin{equation}
\eta_i < 5.81\!\left[1+(1+0.283\,\eta_e)\,\frac{T_{e\,0}}{T_{i\,0}}\right].
\end{equation}

\subsection{Locked Magnetic Islands}\label{slocked}
Consider a locked magnetic island chain: that is, an island chain which is interacting with a static, resonant, external magnetic perturbation whose amplitude is
sufficient to reduce the phase-velocity of the island to zero in the laboratory frame. This implies that $v_p=0$. 

It follows from Eqs.~(\ref{e29h}) and (\ref{e44}) that
\begin{equation}\label{e142}
v= v_{\perp\,i} = 1+\lambda_{\perp\,i}\left(\frac{\eta_i}{1+\eta_i}\right) = \frac{1-1.367\,\eta_i}{1+\eta_i}.
\end{equation}
We conclude that, in the local plasma frame, the phase-velocity of a locked magnetic island chain is solely determined by the  neoclassical
ion perpendicular velocity [which is parameterized by $v_{\perp\,i}$---see Eq.~(\ref{e36t})]. Moreover, the phase-velocity lies between the local perpendicular
guiding-center fluid velocity and the local perpendicular ion fluid velocity (i.e., $0<v<1$)
provided that $0<\eta_i<0.73$. On the other hand, if $\eta_i>0.73$ then the chain rotates in the electron diamagnetic direction (i.e., $v<0$) in the local plasma frame.

The analysis of Sect.~\ref{soln} reveals that the expressions for $J_g$ and $J_b$ are the same for both locked and freely rotating island chains. In other words,
magnetic field-line curvature and the perturbed bootstrap current have the same effect on the stability of a locked island chain as they have on that 
of a corresponding freely rotating chain. 
 On the
other hand, the expression for $J_p$ is [cf., Eq.~(\ref{e138})]
\begin{equation}
J_p=-\left[1.38-Q\left(\frac{\rho_i}{w}\right)\right]\epsilon\,v_{\theta\,i}\,(1-v_{\theta\,i})
\end{equation}
if $\hat{\nu}_{\theta\,i}\gg \hat{\nu}_{\perp\,i}$, and 
\begin{equation}
J_p=-\left[1.38-Q\left(\frac{\rho_i}{w}\right)\right]\epsilon\,v_{\perp\,i}\,(1-v_{\perp\,i})=-\left[1.38-Q\left(\frac{\rho_i}{w}\right)\right]\epsilon\,v\,(1-v)
\end{equation}
if $\hat{\nu}_{\theta\,i}\ll \hat{\nu}_{\perp\,i}$. Here, use has been made of Eq.~(\ref{e142}). It follows that the perturbed ion polarization current has the same effect
on the stability of a locked island chain as it has on that of a corresponding freely rotating chain when the neoclassical ion poloidal flow
damping rate greatly exceeds the neoclassical ion perpendicular flow damping rate (i.e., $\hat{\nu}_{\theta\,i}\gg\hat{\nu}_{\perp\,i}$). On the other hand, if the  neoclassical ion perpendicular flow
damping rate greatly exceeds the neoclassical ion poloidal flow damping rate (i.e., $\hat{\nu}_{\perp\,i}\gg \hat{\nu}_{\theta\,i}$) then the polarization current is stabilizing when $0<\eta_i<0.73$, and destabilizing
otherwise. In the latter case, if $0.73<\eta_i<5.81$ then  the polarization current has a stabilizing effect on a freely rotating island chain, but
a destabilizing effect on a corresponding locked chain. 

\section{Strong Neoclassical Ion Poloidal Flow Damping Regime}\label{s4}
\subsection{Alternative Field Equations}
In the so-called strong neoclassical ion  poloidal flow damping regime, it is helpful to write the field equations (\ref{e11})--(\ref{e15}) in the alternative
form
\begin{align}
0&= [\phi+\tau\,N,\psi]+\beta\,\eta\,J\nonumber\\[0.5ex]
&\phantom{=}+\alpha_n^{-1}\,\hat{\nu}_{\theta\,e}\left[\alpha_n^{-1}\,J +V-\partial_X(\phi+\tau\,v_{\theta\,e}\,N)
-v_{\theta\,i}-\tau\,v_{\theta\,e}\right],\label{e11z}\\[0.5ex]
0&= [\phi,N]-\rho\,[\alpha_n\,V+J,\psi]-\alpha_c\,\rho\,[\phi+\tau\,N,X]+D\,\partial_X^{\,2}N,\\[0.5ex]
0&=\delta\,[\phi,V]-\delta\,\alpha_n\,(1+\tau)\,[N,\psi]+\delta\,\mu\,\partial_X^{\,2}V -\delta\,\hat{\nu}_{\theta\,i}\left[V-\partial_X(\phi-v_{\theta\,i}\,N)\right],\\[0.5ex]
0&=[J,\psi] + \alpha_c\,(1+\tau)\,[N,X] + \partial_X H,\\[0.5ex]
J &= \beta^{\,-1}\left(\partial^{\,2}_X\psi-1\right),
\end{align}
where
\begin{align}
H(X,\zeta) &= \epsilon\,[\phi-N,\partial_X \phi]+[\phi,V]-\alpha_n\,(1+\tau)\,[N,\psi]+\mu\,\partial_X^{\,2}[V+\epsilon\,\partial_X(\phi-N)] \nonumber\\[0.5ex]
&\phantom{=} +\hat{\nu}_{\perp\,i}\left[-\partial_X(\phi-v\,N)\right].\label{e15z}
\end{align}

\subsection{Ordering Scheme}
The ordering scheme adopted in the strong neoclassical ion poloidal flow damping regime is:\,\cite{rf2,rf3}
\begin{tabbing}
\mbox{\hspace{1cm}}\=$\Delta^0\,\delta^{\,-1}$\=:~~~\=$\hat{\nu}_{\theta\,i}$,\\[0.5ex]
\>$\Delta^0\,\delta^{\,0}$\>\>$v_{\theta\,i}$, $v_{\theta\,e}$, $v$, $\tau$, $\alpha_n$, $\alpha_c$\\[0.5ex]
\>$\Delta^0\,\delta^{\,1}$\>: \>$\epsilon$, $\rho$, $\beta$,\\[0.5ex]
\>$\Delta^1\,\delta^{\,0}$\>: \> $\hat{\nu}_{\perp\,i}$, $\eta$, $D$, $\mu$,\\[0.5ex]
\>$\Delta^1\,\delta^{\,1}$\>: \>$\hat{\nu}_{\theta\,e}$.
\end{tabbing}
This ordering scheme is suitable for a constant-$\psi$ (i.e., $\beta\ll 1$) magnetic island chain whose radial width is much larger than the ion
poloidal gyroradius (i.e., $\rho\ll 1$), and which is embedded in a large aspect-ratio (i.e., $\epsilon\ll 1$), high-temperature (i.e., $\eta$, $D$, $\mu\lll 1$)
tokamak plasma equilibrium. The defining feature of the strong neoclassical ion poloidal flow damping regime is that the ion poloidal flow damping rate is sufficiently large that the neoclassical ion stress tensor is the
dominant term in the ion parallel equation of motion. 

\subsection{Order $\Delta^0\,\delta^{\,0}$}
To order $\Delta^0\,\delta^{\,0}$, Eqs.~(\ref{e11z})--(\ref{e15z}) yield
\begin{align}
0&= [\phi_{0,0}+\tau\,N_{0,0},\psi_{0,0}],\label{e53z}\\[0.5ex]
0&= [\phi_{0,0},N_{0,0}],\label{e54z}\\[0.5ex]
0&=-\,\hat{\nu}_{\theta\,i}\left[V_{0,0}-\partial_X(\phi_{0,0}-v_{\theta\,i}\,N_{0,0})\right],\label{e55z}\\[0.5ex]
0&=[J_{0,0},\psi_{0,0}] +\alpha_c\,(1+\tau)\,[N_{0,0},X] +\partial_X\!\left\{[\phi_{0,0},V_{0,0}]-\alpha_n\,(1+\tau)\,[N_{0,0},\psi_{0,0}]\right\},\label{e56z}\\[0.5ex]
\partial_X^{\,2}\psi_{0,0}&=1.\label{e57z}
\end{align}

Equations~(\ref{e39}), (\ref{e45e}), and  (\ref{e57z}) give Eq.~(\ref{epo}).
Equations~(\ref{e40}), (\ref{e41}), (\ref{e53z}), and (\ref{e54z}) lead to  Eqs.~(\ref{e157}) and (\ref{e158}).
 Defining $M({\mit\Omega})$ and $L({\mit\Omega})$ in accordance with Eqs.~(\ref{e59}) and
(\ref{e60}), Eqs.~(\ref{e40}) and (\ref{e41}) yield the boundary conditions (\ref{e61}) and (\ref{e62}). As before, $M=L=0$ inside the separatrix. 
Equations~(\ref{e157})--(\ref{e60}) and (\ref{e55z}) give
\begin{equation}\label{e159}
V_{0,0} = -|X|\,(M-v_{\theta\,i}\,L).
\end{equation}
According to Eqs.~(\ref{e61}) and (\ref{e62}), this expression automatically satisfies the boundary condition (\ref{e42}). 

Equations~(\ref{e157})--(\ref{e60}),  (\ref{e56z}), and (\ref{e159}) yield
\begin{equation}
[J_{0,0},{\mit\Omega}]=\left[\frac{1}{2}\,d_{\mit\Omega}\{(M-v_{\theta\,i}\,L)\,M\}\,X^{\,2}-\alpha_c\,(1+\tau)\,L\,|X|,{\mit\Omega}\right].
\end{equation}
It follows that
\begin{equation}\label{e70xx}
J_{0,0} =\frac{1}{2}\,d_{\mit\Omega}[(M-v_{\theta\,i}\,L)\,M]\,\widetilde{X^{\,2}}-\alpha_c\,(1+\tau)\,L\,\widetilde{|X|} + \bar{J}({\mit\Omega}),
\end{equation}
where $\bar{J}({\mit\Omega})$ is an arbitrary flux function. 
However, the lowest-order flux-surface average of Eq.~(\ref{e11z}) implies that
\begin{equation}\label{e71xx}
\bar{J}({\mit\Omega}) =\alpha_n\left(\frac{\epsilon\,\nu_{\theta\,e}\,\tau_e}{1+\epsilon\,\nu_{\theta\,e}\,\tau_e}\right)(v_{\theta\,i}+\tau\,v_{\theta\,e})\left(1-\frac{L}{\langle 1\rangle}\right),
\end{equation}
where use has been made of Eqs.~(\ref{e157})--(\ref{e60}),   and (\ref{e159}).

Finally, it is easily demonstrated that 
\begin{equation}
X\,[J_{0,0},{\mit\Omega}] = \frac{1}{6}\,[X^{\,3},(M-v_{\theta\,i}\,L)\,M] +\frac{1}{2}\,\alpha_c\,(1+\tau)\,[s\,X^{\,2},N_0],
\end{equation}
which implies that
\begin{equation}
\langle X\,[J_{0,0},{\mit\Omega}]\rangle = 0.
\end{equation}
In other words, $J_{0,0}$ does not contribute to the torque integral, $J_s$ [see Eq.~(\ref{e50})]. 
Thus, in order to calculate $J_s$, and, hence, to determine the phase-velocity of a freely rotating island chain, we must expand to higher order. 

\subsection{Order $\Delta^1\,\delta^{\,0}$}\label{sd10}
To order $\Delta^1\,\delta^{\,0}$, Eqs.~(\ref{epo})--(\ref{e158}), and (\ref{e11z})--(\ref{e15z}) yield
\begin{align}
0&= [\phi_{1,0}+\tau\,N_{1,0},{\mit\Omega}] + s\,[\phi_0+\tau\,N_0,\psi_{1,0}],\label{e75z}\\[0.5ex]
0&= s\,[\phi_{1,0},N_0] +s\,[\phi_0,N_{1,0}]+s\,D\,\partial_X^{\,2} N_0,\label{e76z}\\[0.5ex]
0&=-\hat{\nu}_{\theta\,i}\left[V_{1,0} -\partial_X(\phi_{1,0}-v_{\theta\,i}\,N_{1,0})\right],\label{e77z}\\[0.5ex]
0&=[J_{1,0},{\mit\Omega}]+[J_{0,0},\psi_{1,0}] + \alpha_c\,(1+\tau)\,[N_{1,0},X] + \partial_X H_{1,0},\label{e78z}\\[0.5ex]
\partial_X^{\,2}\psi_{1,0}&=0,\label{e79z}
\end{align}
where
\begin{align}\label{e167z}
H_{1,0} &= [\phi_{1,0},V_{0,0}]+s\,[\phi_0,V_{1,0}]-\alpha_n\,(1+\tau)\,s\,[N_0,\psi_{1,0}]-\alpha_n\,(1+\tau)\,[N_{1,0},{\mit\Omega}]\nonumber\\[0.5ex]
&\phantom{=} +\mu\,\partial_X^{\,2}V_{0,0}+\hat{\nu}_{\perp\,i}\left[-s\,\partial_X(\phi_0-v\,N_0)\right].
\end{align}

It follows from Eq.~(\ref{e79z}) that
\begin{equation}
\psi_{1,0} =0,
\end{equation}
from Eq.~(\ref{e75z}) that
\begin{equation}
\phi_{1,0}=-\tau\,N_{1,0},
\end{equation}
from Eq.~(\ref{e77z}) that
\begin{equation}
V_{1,0} = -(v_{\theta\,i}+\tau)\,\partial_X N_{1,0},
\end{equation}
from Eq.~(\ref{e76z}) that
\begin{equation}\label{e82z}
[N_{1,0},{\mit\Omega}] = D\left(\frac{X^{\,2}\,d_{\mit\Omega}L + L}{M+\tau\,L}\right),
\end{equation}
and from Eqs.~(\ref{e78z}) and (\ref{e167z}) that
\begin{equation}\label{e83z}
[J_{1,0},{\mit\Omega}]=-\alpha_c\,(1+\tau)\,[N_{1,0},X]-\partial_X H_{1,0},
\end{equation}
where 
\begin{align}
H_{1,0 }&= \tau\,[N_{1,0}, |X|\,(M-v_{\theta\,i}\,L)] -(v_{\theta\,i}+\tau)\,M\,[|X|\,d_{\mit\Omega} N_{1,0},{\mit\Omega}]-\alpha_n\,(1+\tau)\,[N_{1,0},{\mit\Omega}]\nonumber\\[0.5ex]
&\phantom{=}-\mu\left[|X|^{\,3}\,d_{\mit\Omega}^{\,2}(M-v_{\theta\,i}\,L)  +3\,|X|\,d_{\mit\Omega}(M-v_{\theta\,i}\,L)\right]
+\hat{\nu}_{\perp\,i}\,|X|\,(M-v\,L).
\end{align}
Here, use has been made of Eqs.~(\ref{epo})--(\ref{e60}), and Eq.~(\ref{e159}). 

As before, the flux-surface average of Eq.~(\ref{e82z}) yields Eq.~(\ref{e85e}).  Equation~(\ref{e82z}) then reduces to 
\begin{equation}
[N_{1,0},{\mit\Omega}] = \left(\frac{D\,d_{\mit\Omega} L}{M+\tau\,L}\right)\widetilde{X^{\,2}}.
\end{equation}
It is clear that $N_{1,0}=0$ inside the island separatrix (because $L=0$ there). Hence, we conclude that $H_{1,0}=0$ inside the separatrix (because $M=L=N_{1,0}$ there).

The flux-surface average of Eq.~(\ref{e83z}) gives
\begin{equation}
d_{\mit\Omega}\!\left\{\langle |X|\,H_{1,0}\rangle +\alpha_c\,(1+\tau)\,\langle |X|\,[N_{1,0},{\mit\Omega}]\rangle\right\}=0,
\end{equation}
which can be integrated to give 
\begin{equation}\label{e179}
\langle |X|\,H_{1,0}\rangle +\alpha_c\,(1+\tau)\,\langle |X|\,[N_{1,0},{\mit\Omega}]\rangle=0.
\end{equation}
Now, it can be demonstrated that\,\cite{rf2}
\begin{align}\label{evvv}
\langle |X|^{\,j}\,H_{1,0}\rangle &= (1+j)^{-1}\left[\tau\,(M-v_{\theta\,i}\,L) -j\,(v_{\theta\,i}+\tau)\,M\right]d_{\mit\Omega}\langle |X|^{\,j+1}\,[N_{1,0},{\mit\Omega}]\rangle\nonumber\\[0.5ex]
&\phantom{=}+\tau\,d_{\mit\Omega}(M-v_{\theta\,i}\,L)\,\langle |X|^{\,j+1}\,[N_{1,0},{\mit\Omega}]\rangle\nonumber\\[0.5ex]
&\phantom{=}+j\,(v_{\theta\,i}+\tau)\,M\,\langle |X|^{\,j-1}\,[N_{1,0},{\mit\Omega}]\rangle-\alpha_n\,(1+\tau)\,\langle |X|^{\,j}\,[N_{1,0},{\mit\Omega}]
\rangle\nonumber\\[0.5ex]
&\phantom{=}-\mu\,\langle |X|^{\,j+3}\rangle\,d_{\mit\Omega}^{\,2}(M-v_{\theta\,i}\,L)-3\,\mu\,\langle |X|^{\,j+1}\rangle\,d_{\mit\Omega}(M-v_{\theta\,i}\,L)\nonumber\\[0.5ex]
&\phantom{=}+\hat{\nu}_{\perp\,i}\,\langle |X|^{\,j+1}\rangle\,(M-v\,L),
\end{align}
where 
\begin{equation}\label{evvv1}
\langle |X|^{\,j}\,[N_{1,0},{\mit\Omega}]\rangle =\left(\frac{D\,d_{\mit\Omega} L}{M+\tau\,L}\right)\left\langle \widetilde{|X|^{\,j}}\,\widetilde{X^{\,2}}\right\rangle,\
\end{equation}
and $j$ is a  non-negative integer. 
Hence, Eq.~(\ref{e179}) yields
\begin{align}\label{e84s}
0&=d_{\mit\Omega}\!\left[\langle X^{\,4}\rangle\,d_{\mit\Omega}(M-v_{\theta\,i}\,L)+\frac{D}{2\,\mu}\,v_{\theta\,i}\left\langle\widetilde{X^{\,2}}\,\widetilde{X^{\,2}}\right\rangle d_{\mit\Omega}L\right]\nonumber\\[0.5ex]
&\phantom{=} -\frac{D}{2\,\mu}\left\langle\widetilde{X^{\,2}}\,\widetilde{X^{\,2}}\right\rangle\left[
(v_{\theta\,i}+2\,\tau)\,d_{\mit\Omega} M- v_{\theta\,i}\,\tau\,d_{\mit\Omega} L\right]\frac{d_{\mit\Omega} L}{M+\tau\,L}\nonumber\\[0.5ex]
&\phantom{=}+\frac{D}{\mu}\,(\alpha_n-\alpha_c)\,(1+\tau)\left\langle\widetilde{|X|}\,\widetilde{X^{\,2}}\right\rangle \frac{d_{\mit\Omega}L}{M+\tau\,L}\nonumber\\[0.5ex]
&\phantom{=}-\frac{\hat{\nu}_{\perp\,i}}{D}\,\langle X^{\,2}\rangle\,(M-v\,L).
\end{align}

\subsection{Evaluation of $J_c$}
According to Eqs.~(\ref{e85e}), (\ref{e70xx}), and (\ref{e71xx}), 
\begin{align}
J_{0,0} &= \frac{1}{2}\,d_{\mit\Omega}\!\left[M\left(M-\frac{v_{\theta\,i}}{\langle X^{\,2}\rangle}\right)\right]\widetilde{X^{\,2}}
-\alpha_c\,(1+\tau)\,\frac{\widetilde{|X|}}{\langle X^{\,2}\rangle}\nonumber\\[0.5ex]
&\phantom{=} +\alpha_n\left(\frac{\epsilon\,\nu_{\theta\,e}\,\tau_e}{1+\epsilon\,\nu_{\theta\,e}\,\tau_e}\right)(v_{\theta\,i}+\tau\,v_{\theta\,e})\left(1-\frac{1}{\langle 1\rangle\langle X^{\,2}\rangle}\right)
\end{align}
for ${\mit\Omega}\geq 1$, and
\begin{equation}
J_{0,0} = \alpha_n\left(\frac{\epsilon\,\nu_{\theta\,e}\,\tau_e}{1+\epsilon\,\nu_{\theta\,e}\,\tau_e}\right)(v_{\theta\,i}+\tau\,v_{\theta\,e})
\end{equation}
for $-1\leq {\mit\Omega}< 1$. 
Thus, it follows from Eq.~(\ref{e49})  that
$J_c = J_p + J_g+ J_b$,
where
\begin{equation}\label{ejp}
J_p =\int_{1-}^{\infty} d_{\mit\Omega}\!\left[M\left(M-\frac{v_{\theta\,i}}{\langle X^{\,2}\rangle}\right)\right]\langle \widetilde{X^{\,2}}\,\widetilde{X^{\,2}}\rangle\,d{\mit\Omega}
\end{equation}
parameterizes the effect of the perturbed ion polarization current on island stability,  whereas 
\begin{equation}
J_g = -\alpha_c\,(1+\tau)\int_1^{\infty}2\,\frac{\langle\widetilde{|X|}\,\widetilde{X^{\,2}}\rangle}{\langle X^{\,2}\rangle}\,d{\mit\Omega}=-I_3\,\alpha_c\,(1+\tau)
\end{equation}
(see the Appendix for the definition of $I_3=1.58$) parameterizes the effect of magnetic field-line curvature on island stability, and, finally, 
\begin{align}
J_b &=-\alpha_n\left(\frac{\epsilon\,\nu_{\theta\,e}\,\tau_e}{1+\epsilon\,\nu_{\theta\,e}\,\tau_e}\right)\,(v_{\theta\,i}+v_{\theta\,e}) \int_1^{\infty}2\,\frac{\langle\widetilde{|X|}\,\widetilde{X^{\,2}}\rangle}{\langle X^{\,2}\rangle}\,d{\mit\Omega}\nonumber\\[0.5ex]
&=-I_3\,\alpha_n\left(\frac{\epsilon\,\nu_{\theta\,e}\,\tau_e}{1+\epsilon\,\nu_{\theta\,e}\,\tau_e}\right)\,(v_{\theta\,i}+v_{\theta\,e}) 
\end{align}
parameterizes the effect of the perturbed bootstrap current on island stability. It can be seen, by comparison with the analysis of Sect.~\ref{soln},
that $J_g$ has the same form in both the weak and the strong  neoclassical ion flow damping regimes, whereas $J_b$ has very
similar forms in the two regimes (in fact, the forms are identical if  $\epsilon\,\nu_{\theta\,e}\,\tau_e=1.67\,f_t=2.44\,\epsilon_s^{\,1/2}\ll 1$). 

\subsection{Evaluation of $J_s$}
Multiplying Eq.~(\ref{e83z}) by $X$, and flux-surface averaging, we obtain
\begin{equation}
\langle X\,[J_{0,1},{\mit\Omega}]\rangle = -d_{\mit\Omega}\!\left\{\langle X^{\,2}\,H_{1,0}\rangle-\frac{\alpha_c}{2}\,(1+\tau)\,\langle X^{\,2}\,[N_{1,0},{\mit\Omega}]\rangle\right\}
+\langle H_{1,0}\rangle,
\end{equation}
which can be integrated to give
\begin{equation}
\int_{-1}^\infty \langle X\,[J_{0,1},{\mit\Omega}]\rangle \,d{\mit\Omega} = \int_1^\infty \langle H_{1,0}\rangle\,d{\mit\Omega},
\end{equation}
because $H_{1,0}=0$ inside the separatrix. 
Making use of Eqs.~(\ref{e50}), (\ref{evvv}), and (\ref{evvv1}), we obtain
\begin{equation}\label{ejs}
J_s = -\hat{\nu}_{\perp\,i}\int_1^\infty 4\left(M-v\,L\right)d{\mit\Omega}.
\end{equation}

\subsection{Separatrix Boundary Layer}\label{sbound}
The flux-surface functions $M({\mit\Omega})$ and $L({\mit\Omega})$ are both zero inside, and non-zero just outside, the magnetic separatrix. 
Retaining selected higher-order terms (containing radial derivatives) in Eqs.~(\ref{e76z}) and (\ref{e78z}), 
 we find that
\begin{align}
(M+\tau\,L)\,[N_{1,0},{\mit\Omega}]-s\,\rho\,[J_{1,0},{\mit\Omega}] &\simeq D\left(X^{\,2}\,d_{\mit\Omega} L + L\right),\\[0.5ex]
s\,[J_{1,0},{\mit\Omega}] &\simeq -\partial_X[\phi_0, V_{1,0}]=(v_{\theta\,i}+\tau)\,M\,\partial_X^{\,2}[N_{1,0},{\mit\Omega}],
\end{align}
so that Eq.~(\ref{e82z}) generalizes to give
\begin{equation}
\left\{(M+\tau\,L)-(v_{\theta\,i}+\tau)\,M\,\rho\,\partial_X^{\,2}\right\}[N_{1,0},{\mit\Omega}]\simeq  D\left(X^{\,2}\,d_{\mit\Omega} L + L\right),
\end{equation}
which suggests that the apparent discontinuities in the functions $M({\mit\Omega})$ and $L({\mit\Omega})$ are
resolved in a thin boundary layer of (unnormalized) width $(\rho)^{1/2}\,w=\rho_{\theta\,i}$ on the island separatrix.

 Inside the boundary layer, Eq.~(\ref{e84s}) reduces to
\begin{equation}
0\simeq d_y^{\,2}\!\left(M-v_{\theta\,i}\,L+\frac{D}{2\,\mu}\,v_{\theta\,i}\,L\right)-\frac{D}{2\,\mu}\,d_y\!\left\{(v_{\theta\,i}+2\,\tau)\,M-v_{\theta\,i}\,\tau\,L\right\}\frac{d_y L}{M+\tau\,L},
\end{equation}
where $y=({\mit\Omega}-1)/(\rho_{\theta\,i}/w)$, and $d_y\equiv d/dy$.  Note that $d_{\mit\Omega}\sim {\cal O}(w/\rho_{\theta\,i})\,d_y\gg 1$. Here, we
have made use of the fact that $\langle X^4\rangle =\langle\widetilde{X^2}\,\widetilde{X^2}\rangle$ close to the separatrix. 
Let us assume that $M=v_{\theta\,i}\,(1-v_0)\,L$ within the layer, where $v_0$ is a constant. It follows that
\begin{equation}
0\simeq d_y(L\,d_y L) - \left[\frac{D}{2\,\mu}\,\frac{(1+2\,\hat{\tau})\,(1-v_0)-\hat{\tau}}{(1-v_0+\hat{\tau})\,(v_0-D/2\,\mu)}-1\right]
(d_y L)^{\,2},
\end{equation}
where $\hat{\tau}=\tau/v_{\theta\,i}$. 
Integrating across the layer from just inside the separatrix ({\rm i.e.}, $y\rightarrow-\infty$, where $L=0$) to
just outside the separatrix [{\rm i.e.}, $y\rightarrow\infty$, where $d_y L\sim {\cal O}(\rho_{\theta\,i}/w)\ll 1$, because  $d_{\mit\Omega} L\sim {\cal O}(1)$], we obtain
\begin{equation}\label{e79x}
 \left[\frac{D}{2\,\mu}\,\frac{(1+2\,\hat{\tau})\,(1-v_0)-\hat{\tau}}{(1-v_0+\hat{\tau})\,(v_0-D/2\,\mu)}-1\right]
\int_{-\infty}^\infty (d_y L)^{\,2}\,dy\ll 1.
\end{equation}
Now, the integral in the previous expression is positive definite, and also of order unity. Thus, the only way in which Eq.~(\ref{e79x}) can
be satisfied is if
\begin{equation}
\frac{D}{2\,\mu}\,\frac{(1+2\,\hat{\tau})\,(1-v_0)-\hat{\tau}}{(1-v_0+\hat{\tau})\,(v_0-D/2\,\mu)}= 1, 
\end{equation}
or
\begin{equation}\label{e89f}
v_0^{\,2} -(1+\hat{\tau})\left(1+\frac{D}{\mu}\right) v_0+(1+\hat{\tau})\,\frac{D}{\mu}= 0,
\end{equation}
which implies that
\begin{equation}\label{ev0}
v_0 = \left(\frac{1+\hat{\tau}}{2}\right)\left(1+\frac{D}{\mu}-\left[1-2\,\frac{D}{\mu}\left(\frac{1-\hat{\tau}}{1+\hat{\tau}}\right)+\left(\frac{D}{\mu}\right)^2\right]^{1/2}\right).
\end{equation}
Here, we have chosen the root of the quadratic equation (\ref{e89f}) that corresponds to the obvious physical solution $v_0=0$ when $D/\mu=0$.\cite{hyp1} Note that $0\leq v_0\leq 1$. 

\subsection{Transformed Equations}
Equation~(\ref{e84s}) reduces to 
\begin{align}\label{e194}
0&=d_k\!\left[2\,k^{\,2}\,{\cal E}\,d_k M + v_{\theta\,i}\left(1-\frac{D}{2\,\mu}\right)\left(\frac{{\cal E}\,{\cal A}}{{\cal C}^{\,2}}-1\right)\right]\nonumber\\[0.5ex]
&\phantom{=} + \frac{D}{2\,\mu}\left(\frac{{\cal E}\,{\cal A}}{{\cal C}^{\,2}}-1\right)\left[
\frac{(v_{\theta\,i}+2\,\tau)\,2\,k\,{\cal C}\,d_k M + v_{\theta\,i}\,\tau\,{\cal A}/k\,{\cal C}}{2\,k\,{\cal C}\,M+\tau}\right]\nonumber\\[0.5ex]
&\phantom{=}-\frac{D}{\mu}\,(\alpha_n-\alpha_c)\,(1+\tau)\left(\frac{{\cal D}\,{\cal A}}{{\cal C}}-1\right)\left(\frac{4\,k}{2\,k\,{\cal C}\,M+\tau}\right)\nonumber\\[0.5ex]
&\phantom{=}-\frac{\hat{\nu}_{\perp\,i}}{\mu}\,4\,k\,(2\,k\,{\cal C}\,M - v),
\end{align}
where $k=[(1+{\mit\Omega})/2]^{\,1/2}$ and ${\cal A}(k)$, ${\cal C}(k)$, ${\cal D}(k)$, and ${\cal E}(k)$ are defined in the Appendix. It follows from Eq.~(\ref{e61}), and the analysis
of Sect.~\ref{sbound}, that
\begin{align}\label{e195}
M(1) &= v_{\theta\,i}\,(1-v_0)\,\frac{\pi}{4},\\[0.5ex]
M(k\rightarrow\infty) &= \frac{v}{2\,k}.\label{e196}
\end{align}
Furthermore, Equations~(\ref{ejp}) and (\ref{ejs}) yield
\begin{align}\label{e198}
J_p &=-\left[\frac{2\pi}{3}-Q\left(\frac{\rho_{\theta\,i}}{w}\right)\right]v_{\theta\,i}^{\,2}\,v_0\,(1-v_0)
+\int_{1+}^\infty d_k\!\left[M\left(M-\frac{v_{\theta\,i}}{2\,k\,{\cal C}}\right)\right]8\left({\cal E}-\frac{{\cal C}^{\,2}}{{\cal A}}\right)k^{\,3}\,dk,\\[0.5ex]
J_s &= -\hat{\nu}_{\perp\,i}\int_1^\infty 16\left(M\,k-\frac{v}{2\,{\cal C}}\right)dk,\label{e199}
\end{align}
respectively. Here, we have evaluated the contribution to the polarization integral emanating from the boundary layer on the magnetic
separatrix [i.e., the first term on the right-hand side of Eq.~(\ref{e198})] according to the method set out in Sect.~\ref{sq}, taking into account
the fact that that the (unnormalized) thickness of the layer is $\rho_{\theta\,i}$. 
 It remains to solve Eq.~(\ref{e194}), subject to the boundary conditions (\ref{e195}) and (\ref{e196}), and then to evaluate the integrals (\ref{e198}) and (\ref{e199}). This task, which involves the elimination of an unphysical solution that varies as $M\sim \exp[2\,(\hat{\nu}_{\perp\,i}/\mu)^{1/2}\,k]$ at
large $k$, can be performed analytically in the strong ion perpendicular flow damping regime, $\hat{\nu}_{\perp\,i}\gg \mu$, but must, otherwise,
be performed numerically. 

\subsection{Weak Neoclassical Ion Perpendicular Flow Damping Regime}\label{www}
Suppose that $\hat{\nu}_{\perp\,i}/\mu\ll 1$. In the limit $k\gg 1$, Eq.~(\ref{e194}) reduces to
\begin{equation}
d_k(2\,k^{\,2}\,d_k M) -\frac{\hat{\nu}_{\perp\,i}}{\mu}\,4\,k\,(2\,k\,M - v)= 0.
\end{equation}
The solution is
\begin{equation}\label{e200}
M(k)= \frac{v+(v_{\theta\,i}\,v_f-v)\,{\rm e}^{-2\,(\hat{\nu}_{\perp\,i}/\mu)^{1/2}\,k}}{2\,k},
\end{equation}
where use has been made of Eq.~(\ref{e196}). Here, $v_f$ is an arbitrary constant. 

It follows from Eq.~(\ref{e199}) that 
\begin{equation}
J_s= -4\,(\hat{\nu}_{\perp\,i}\,\mu)^{1/2}\,(v_{\theta\,i}\,v_f-v).
\end{equation}
Hence, $v_f$ determines the phase-velocity of a freely rotating magnetic island chain. (See Sect.~\ref{sfree}.)

\begin{figure}
\epsfysize=4in
\centerline{\epsffile{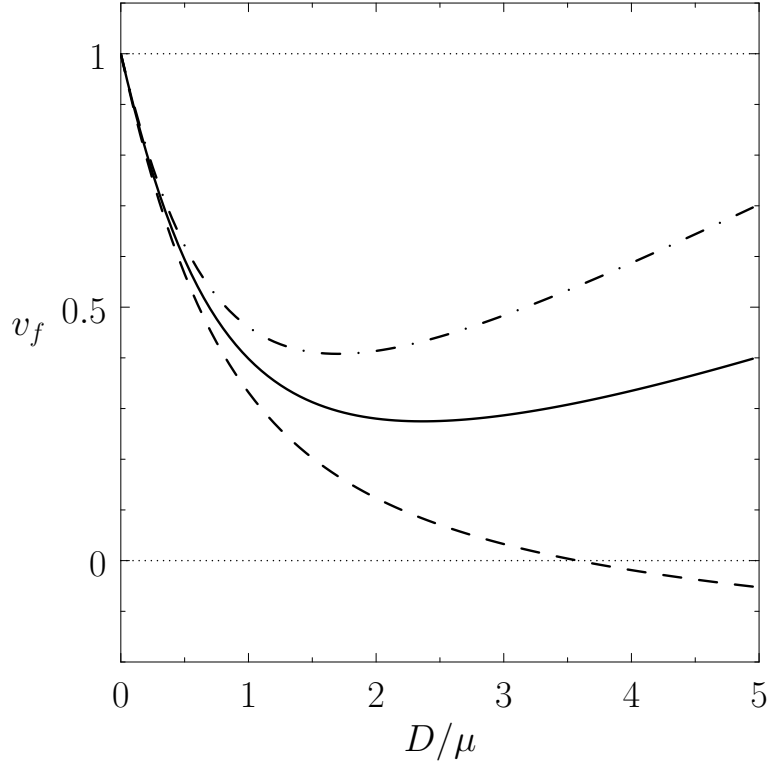}}
\caption{The island phase-velocity parameter, $v_f$, calculated as a function of the perpendicular diffusivity ratio, $D/\mu$, in the
weak neoclassical ion perpendicular flow damping limit of the strong neoclassical ion poloidal flow damping regime. 
The dashed, solid, and dash-dotted curves correspond to ${\mit\Lambda} = 0.5$, $0.0$, and $-0.5$, respectively. All
curves are calculated with $\hat{\tau}=1$ and $\rho_{\theta\,i}/w=1\times 10^{-3}$. }\label{f2}
\end{figure}

\begin{figure}
\epsfysize=4in
\centerline{\epsffile{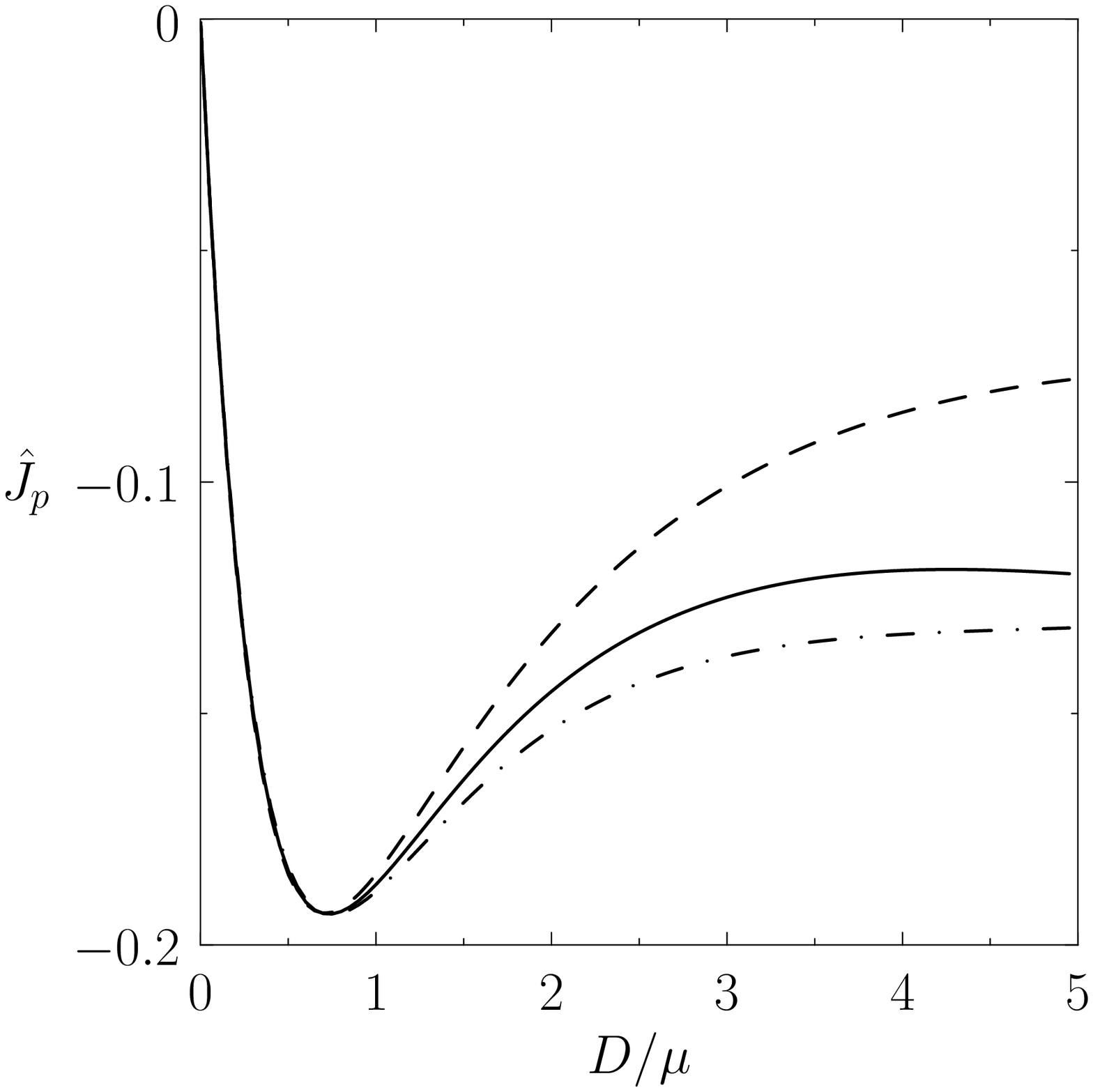}}
\caption{The normalized ion polarization current integral, $\hat{J}_p$, calculated as a function of the perpendicular diffusivity ratio, $D/\mu$, in the
weak neoclassical ion perpendicular flow damping limit of the strong neoclassical ion poloidal flow damping regime. The dashed, solid, and dash-dotted curves correspond to ${\mit\Lambda} = 0.5$, $0.0$, and $-0.5$, respectively. All
curves are calculated with $\hat{\tau}=1$ and $\rho_{\theta\,i}/w=10^{-3}$. }\label{f3}
\end{figure}

\begin{figure}
\epsfysize=4in
\centerline{\epsffile{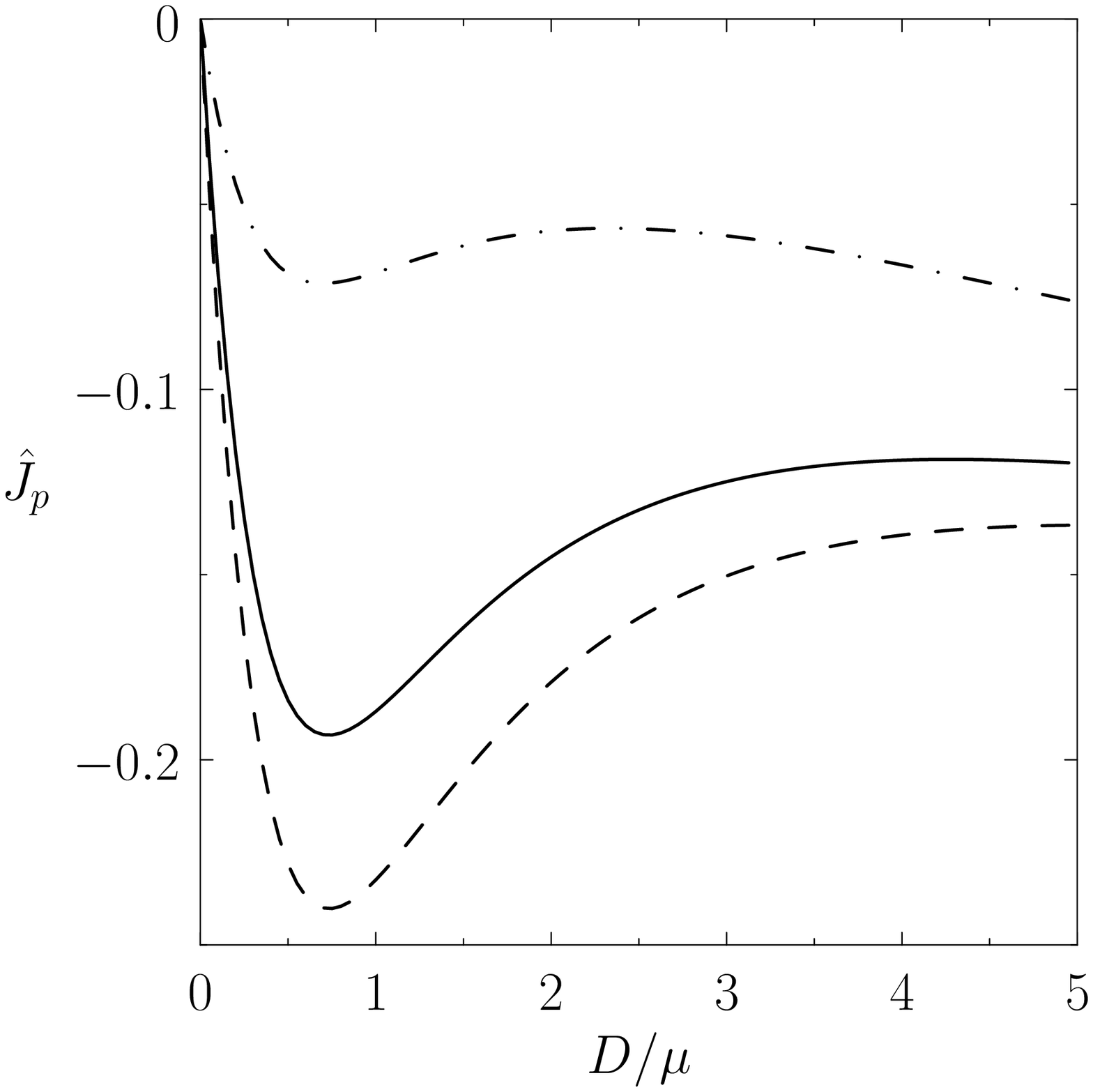}}
\caption{The normalized ion polarization current integral, $\hat{J}_p$,  calculated as a function of the perpendicular diffusivity ratio, $D/\mu$, in the
weak neoclassical ion perpendicular flow damping limit of the strong neoclassical ion poloidal flow damping regime. The dashed, solid, and dash-dotted curves correspond to $\rho_{\theta\,i}/w = 10^{-5}$, $10^{-3}$, and $10^{-1}$, respectively. All
curves are calculated with $\hat{\tau}=1$ and ${\mit\Lambda}=0$. }\label{f3a}
\end{figure}

In the region $1\leq  k\ll (\mu/\hat{\nu}_{\perp\,i})^{\,1/2}$, Eq.~(\ref{e194}) reduces to 
\begin{align}\label{e202}
0&=d_k\!\left[2\,k^{\,2}\,{\cal E}\,d_k \widehat{M} + \left(1-\frac{D}{2\,\mu}\right)\frac{{\cal E}\,{\cal A}}{{\cal C}^{\,2}}\right]\nonumber\\[0.5ex]
&\phantom{=} + \frac{D}{2\,\mu}\left(\frac{{\cal E}\,{\cal A}}{{\cal C}^{\,2}}-1\right)\left[
\frac{(1+2\,\hat{\tau})\,2\,k\,{\cal C}\,d_k \widehat{M} + \hat{\tau}\,{\cal A}/k\,{\cal C}}{2\,k\,{\cal C}\,\widehat{M}+\hat{\tau}}\right]\nonumber\\[0.5ex]
&\phantom{=}-\frac{D}{\mu}\,{\mit\Lambda}\left(\frac{{\cal D}\,{\cal A}}{{\cal C}}-1\right)\left[\frac{4\,k\,(1+\hat{\tau})}{2\,k\,{\cal C}\,\widehat{M}+\hat{\tau}}\right],
\end{align}
where
\begin{align}
\widehat{M}(k) &= \frac{M(k)}{v_{\theta\,i}},\\[0.5ex]
{\mit\Lambda} &= \frac{(\alpha_n-\alpha_c)\,(1+\tau)}{v_{\theta\,i}\,(v_{\theta\,i}+\tau)}.
\end{align}
Recall that $\hat{\tau}=\tau/v_{\theta\,i}$. 
Equation~(\ref{e202}) must be solved subject to the boundary conditions
\begin{align}\label{e204}
\widehat{M}(1) &= (1-v_0)\,\frac{\pi}{4},\\[0.5ex]
\widehat{M}(k\rightarrow\infty) &= \frac{v_f}{2\,k},\label{e205}
\end{align}
where use has been made of Eqs.~(\ref{e195}) and (\ref{e200}). Finally, 
\begin{equation}
J_p = v_{\theta\,i}^{\,2}\,\hat{J}_p,
\end{equation}
where
\begin{equation}\label{e208}
\hat{J}_p =-\left[\frac{2\pi}{3}-Q\left(\frac{\rho_{\theta\,i}}{w}\right)\right]v_0\,(1-v_0)
+\int_{1+}^\infty d_k\!\left[\widehat{M}\left(\widehat{M}-\frac{1}{2\,k\,{\cal C}}\right)\right]8\left({\cal E}-\frac{{\cal C}^{\,2}}{{\cal A}}\right)k^{\,3}\,dk.
\end{equation}
It is clear that $v_f=v_f(D/\mu,{\mit\Lambda},\hat{\tau})$ and $\hat{J}_p=\hat{J}_p(D/\mu,{\mit\Lambda},\hat{\tau},\rho_{\theta\,i}/w)$.
Figures~\ref{f2} and \ref{f3} show $v_f$ and $\hat{J}_p$, calculated numerically as functions of $D/\mu$,  for 
$\hat{\tau}=1$, $\rho_{\theta\,i}/w=10^{-3}$,  and ${\mit\Lambda} = 0.5$, $0.0$, and $-0.5$.  Note that $v<1$, which implies that a freely rotating island chain propagates in the
electron diamagnetic direction relative to the unperturbed local ion fluid, and $\hat{J}_p<0$, which implies that the perturbed polarization current has
a stabilizing effect on the chain. Figure~\ref{f3a} shows $\hat{J}_p$, calculated as a function of $D/\mu$,  for 
$\hat{\tau}=1$, ${\mit\Lambda}=0$,  and $\rho_{\theta\,i}/w= 10^{-5}$, $10^{-3}$, and $10^{-1}$.  It can be seen that the magnitude of the polarization
integral, $\hat{J}_p$,  decreases significantly as the relative 
width of the separatrix boundary layer increases.\cite{james1,james2} However, the sign of the integral remains negative. This shows that, although the contribution
of the separatrix boundary layer to the polarization integral is reduced when the finite width of the layer is taken into account, the layer contribution
still remains large enough to determine the sign of the integral. (Note that if the layer contribution were entirely neglected then the integral would be positive.)

\subsection{Strong Neoclassical Ion Perpendicular Flow Damping Regime}\label{sss}
Suppose that $\hat{\nu}_{\perp\,i}/\mu\gg 1$. Equations~(\ref{e194})--(\ref{e196}) give 
\begin{equation}
M(k) = \frac{v+[v_{\theta\,i}\,(1-v_0)-v]\,{\rm e}^{-(6\,\hat{\nu}_{\perp\,i}/\mu)^{1/2}\,(k-1)}}{2\,k\,{\cal C}}.
\end{equation}
It follows from Eqs.~(\ref{e198}) and (\ref{e199}) that 
\begin{align}\label{e210}
J_p &=  -\left[I_2-Q\left(\frac{\delta_\perp}{w}\right)\right]v\,(v_{\theta\,i}-v)
-\left[Q\left(\frac{\delta_\perp}{w}\right)-Q\left(\frac{\rho_{\theta\,i}}{w}\right)\right]v_{\theta\,i}^{\,2}\,v_0\,(1-v_0),
\end{align}
and
\begin{equation}\label{e211}
J_s = - \sqrt{\frac{8}{3}}\,\pi\,(\hat{\nu}_{\perp\,i}\,\mu)^{\,1/2}\left[v_{\theta\,i}\,(1-v_0)-v\right],
\end{equation}
respectively.  Here, 
\begin{equation}
\delta_\perp =\left( \frac{q_s}{\epsilon_0}\right)\left(\frac{2}{3}\,\frac{\mu_{\perp\,i}}{n_0\,m_i\,\nu_{\perp\,i}}\right)^{1/2},
\end{equation}
and it is assumed that $\rho_{\theta\,i}\ll \delta_\perp\ll w$. 
Note that we have again evaluated the contributions to the polarization integral emanating from  boundary layers on the magnetic
separatrix [i.e., both terms on the right-hand side of Eq.~(\ref{e210})] according to the method set out in Sect.~\ref{sq}.
Furthermore, $I_2=1.38$ is defined in the Appendix.

\subsection{Freely Rotating Magnetic Islands}\label{sfree1}
This subsection, and the following subsection, will concentrate on the strong neoclassical perpendicular flow damping regime, discussed in Sect.~\ref{sss}, 
for which we possess an analytic solution. 

Consider a freely rotating magnetic island chain. As discussed in Sect.~\ref{sfree},  there is zero local drag torque acting on such a chain  (i.e., $J_s=0$). Thus,
it follows from Eqs.~(\ref{e28h}), (\ref{ev0}), and (\ref{e211}) that the chain's phase-velocity parameter is given
by
\begin{equation}
v= v_{\theta\,i}\,(1-v_0) =\left(\frac{v_{\theta\,i}+\tau}{2}\right)\left(\frac{v_{\theta\,i}+\tau}{v_{\theta\,i}-\tau}-\frac{D}{\mu}
+\left[1-2\,\frac{D}{\mu}\left(\frac{v_{\theta\,i}+\tau}{v_{\theta\,i}-\tau}\right)+\left(\frac{D}{\mu}\right)^2\right]^{1/2}\right),
\end{equation}
where 
\begin{equation}
v_{\theta\,i} = \frac{1-0.172\,\eta_i}{1+\eta_i}.
\end{equation}
In the limit $D/\mu\ll 1$, the previous two equations reduce to
\begin{equation}
v\simeq  \left(\frac{1-0.172\,\eta_i}{1+\eta_i}\right)\left(1-\frac{D}{\mu}\right),
\end{equation}
whereas in the opposite limit $\mu/D\ll 1$, we get
\begin{equation}
v\simeq \left(\frac{1-0.172\,\eta_i}{1+\eta_i}\right)\frac{\mu}{D}.
\end{equation}
We conclude that the phase-velocity of a freely rotating magnetic island chain is determined by the  neoclassical
ion poloidal velocity (which is parameterized by $v_{\theta\,i}$), the ratio of the perpendicular particle and momentum diffusivities (which is parameterized by $D/\mu$),
and the electron-ion temperature ratio (which is parameterized by $\tau$). 
The phase-velocity   lies between the unperturbed local perpendicular guiding-center fluid velocity and the unperturbed local
perpendicular ion fluid velocity (i.e., $0<v<1$, as seen in experiments\,\cite{lhaye})  provided that $0<\eta_i<5.81$. On the other hand, if $\eta_i>5.81$ then the chain rotates in the electron diamagnetic
direction (i.e., $v<0$). 

According to Eq.~(\ref{e210}), 
\begin{equation}
J_p =-\left[1.38-Q\left(\frac{\rho_{\theta\,i}}{w}\right)\right]v_{\theta\,i}^{\,2}\,v_0\,(1-v_0),
\end{equation}
where $v_0$ is specified in Eq.~(\ref{ev0}). Note that $0<v_0< 1$. In the limit $D/\mu\ll 1$, we get
\begin{equation}
J_p \simeq -\left[1.38-Q\left(\frac{\rho_{\theta\,i}}{w}\right)\right]\left( \frac{1-0.172\,\eta_i}{1+\eta_i}\right)^2\frac{D}{\mu},
\end{equation}
whereas in the opposite limit $\mu/D\ll 1$, we obtain
\begin{equation}
J_p\simeq -\left[1.38-Q\left(\frac{\rho_{\theta\,i}}{w}\right)\right]\left(\frac{1-0.172\,\eta_i}{1+\eta_i}\right)^2\frac{\mu}{D}.
\end{equation}
Assuming that $1.38> Q(\rho_{\theta\,i}/w)$ (which must be the case, otherwise the width of the separatrix boundary layer would
be comparable with that of the island, thus, invalidating our analysis---see Fig.~\ref{f0}), we conclude that the perturbed ion polarization current always has a stabilizing effect on the island chain (i.e. $J_p<0$). 
Note, incidentally, that, in the strong neoclassical ion poloidal flow
damping regime, $J_p$ is  a factor $\epsilon^{\,-1}=(q_s/\epsilon_s)^{\,2}$ larger  in magnitude than  in the weak neoclassical ion poloidal flow
damping regime (see Sect.~\ref{sfree}).\cite{rob,re} 

\subsection{Locked Magnetic Islands}\label{slocked1}
Consider a locked magnetic island chain, which is characterized by $v_p=0$. It follows from Eqs.~(\ref{e29h}) and (\ref{e44}) that
\begin{equation}
v= v_{\perp\,i} = 1+\lambda_{\perp\,i}\left(\frac{\eta_i}{1+\eta_i}\right) = \frac{1-1.367\,\eta_i}{1+\eta_i},
\end{equation}
We conclude that, in the local plasma frame, the phase-velocity of a locked magnetic island  chain is solely determined by the  neoclassical
ion perpendicular velocity (which is parameterized by $v_{\perp\,i}$). Moreover, the phase-velocity lies between the local perpendicular
guiding-center fluid velocity and the local perpendicular ion fluid velocity (i.e., $0<v<1$)
provided that $0<\eta_i<0.73$. On the other hand, if $\eta_i>0.73$ then the chain rotates in the electron diamagnetic direction (i.e., $v<0$) in the local plasma frame.

\begin{figure}
\epsfysize=4in
\centerline{\epsffile{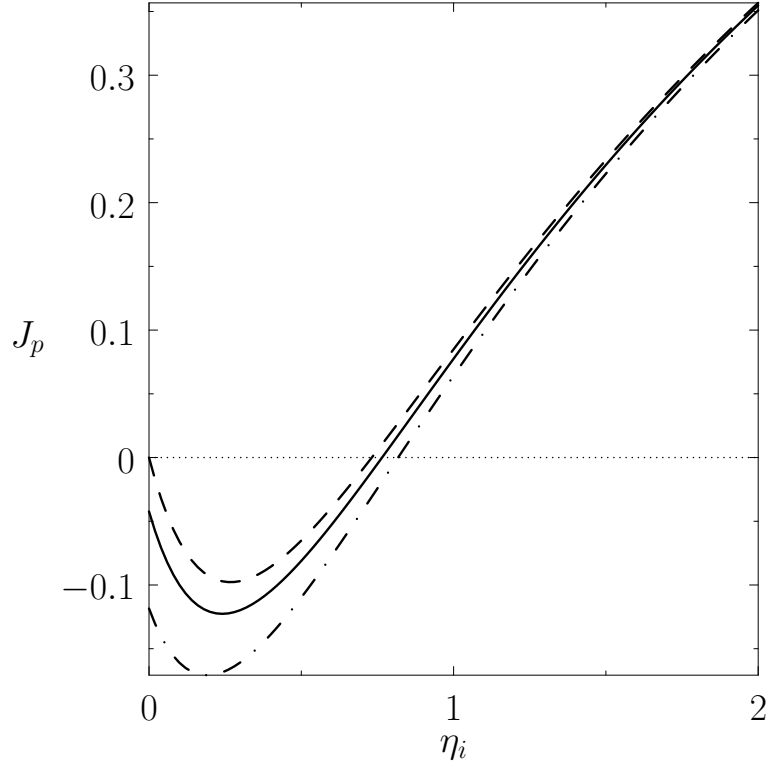}}
\caption{The ion polarization current integral, $J_p$, calculated as a function of the ion temperature gradient parameter, $\eta_i$, for a locked island in the
strong neoclassical ion perpendicular flow damping limit of the strong neoclassical ion poloidal flow damping regime. The dashed, solid, and dash-dotted curves correspond to $\delta_\perp/w = 10^{-3}$, $10^{-2}$, and $10^{-1}$, respectively. All
curves are calculated with $\hat{\tau}=1$, $D/\mu=1$, and $\rho_{\theta\,i}/w=10^{-3}$. }\label{f3b}
\end{figure}

According to Eqs.~(\ref{e28h}), (\ref{e29h}),   (\ref{e44}), and (\ref{e210}), 
\begin{align}
J_p&= -\left[1.38-Q\left(\frac{\rho_{\theta\,i}}{w}\right)\right]v_{\perp\,i}\,(v_{\theta\,i}-v_{\perp\,i})-\left[Q\left(\frac{\delta_\perp}{w}\right)-Q\left(\frac{\rho_{\theta\,i}}{w}\right)\right]v_{\theta\,i}^{\,2}\,v_0\,(1-v_0) \nonumber\\[0.5ex]
&= -1.65\,\frac{\eta_i\,(1-1.367\,\eta_i)}{(1+\eta_i)^{\,2}}\left[1 - 0.72\,Q\left(\frac{\rho_{\theta\,i}}{w}\right)\right]\nonumber\\[0.5ex]
&\phantom{=}-
\left(\frac{1-0.172\,\eta_i}{1+\eta_i}\right)^{\,2}v_0\,(1-v_0)\left[Q\left(\frac{\delta_\perp}{w}\right)-Q\left(\frac{\rho_{\theta\,i}}{w}\right)\right].
\end{align}
Figure~\ref{f3b} shows $J_p$ plotted as a function of $\eta_i$ for various different values of $\delta_\perp/\rho_{\theta\,i}$. 
It can be seen that the perturbed ion polarization current has a stabilizing effect on a locked magnetic island  chain (i.e., $J_p<0$) when $\eta_i$ is less than about $0.75$, and a
destabilizing effect when it exceeds this value. 

\section{Summary and Conclusions}\label{s5}
In this paper, we have calculated the effect of the perturbed ion polarization current on the stability of ion-branch, 
neoclassical tearing modes using an improved,  neoclassical, four-field, drift-MHD model.  
The improvements to the model are described in the Introduction.
The calculation involves the self-consistent
determination of the pressure and scalar electric potential profiles in the vicinity of the associated magnetic island chain, which allows a
determination of the  chain's propagation velocity. We have considered two regimes. First, the so-called {\em weak neoclassical ion poloidal  flow damping regime}\/ 
in which  neoclassical ion poloidal flow damping is not strong enough to enhance the magnitude of the polarization
current (relative to that found in slab geometry)---see Sect.~\ref{s3}. Second, the so-called {\em strong neoclassical ion poloidal flow damping regime}\/  in which neoclassical
ion poloidal flow damping is strong enough to significantly enhance the magnitude of the polarization current---see Sect.~\ref{s4}. 
In both regimes, we have considered two types of solution. First,  {\em freely rotating}\/ solutions (i.e., island chains that
are not interacting with  static, resonant, magnetic perturbations)---see Sects.~\ref{sfree} and \ref{sfree1}. 
Second,  {\em locked}\/ solutions (i.e., island chains that have been brought to rest in the laboratory frame via interaction
with  static, resonant, magnetic perturbations)---see Sects.~\ref{slocked} and \ref{slocked1}. 

In the weak neoclassical ion poloidal flow damping regime,  the island width evolution equation of a freely rotating island
chain takes the form 
\begin{align}\label{ew}
0.823\,\tau_R\,\frac{d}{dt}\!\left(\frac{W}{r_s}\right)& = {\mit\Delta}'\,r_s \nonumber\\[0.5ex]
&\phantom{=}+15.41\,\epsilon_s^{\,1/2}\,\beta_p\left(\frac{L_q}{L_n}\right)\left(\frac{r_s}{W}\right)
\left[(1-0.172\,\eta_i)+(1+0.283\,\eta_e)\,\frac{T_{e}}{T_{i}}\right]\nonumber\\[0.5ex]
&\phantom{=}-12.64\,\beta_p\left(\frac{L_q^{\,2}}{L_n\,L_c}\right)\left(\frac{r_s}{W}\right)\left[(1+\eta_i)+(1+\eta_i)\,\frac{T_e}{T_i}\right]\nonumber\\[0.5ex]
&\phantom{=} -103.5\,\beta_p\left(\frac{L_q}{L_n}\right)^{\,2}\left(\frac{\rho_i}{W}\right)^{\,2}\left(\frac{r_s}{W}\right)P\left(\frac{\rho_i}{W}\right)\,\eta_i\,(1-0.172\,\eta_i),
\end{align}
where
\begin{equation}\label{pdef}
P(x) = 1-\frac{4.5}{\ln(4/x)}-\frac{2.2}{\ln^{\,2}(4/x)}.
\end{equation}
Here, $W=4\,w$ is the full radial island width, $r_s$ the minor radius of the rational surface, $\epsilon_s=r_s/R_0$ the inverse aspect-ratio, $R_0$ the major radius, $\tau_R=\mu_0\,r_s^{\,2}/\eta_{\parallel}$ the
resistive diffusion timescale, $\eta_\parallel$ the parallel resistivity, ${\mit\Delta}'$ the tearing stability index, $\beta_p=\mu_0\,n_e\,T_{i}/B_\theta^{\,2}$ the poloidal ion beta,
$\rho_i = (T_{i}/m_i)^{1/2}\,(m_i/e\,B_\varphi)$ the ion gyroradius, $B_\theta$ the equilibrium poloidal magnetic field-strength, $B_\varphi$ the
equilibrium toroidal magnetic field-strength, $m_i$ the ion mass, $e$ the magnitude of the electron charge, $n_e$ the equilibrium electron number density, $T_i$ the equilibrium  ion temperature,
$T_e$ the equilibrium electron temperature, $L_n$ the equilibrium density scale-length, $L_q$ the equilibrium safety-factor scale-length, $L_c$ the mean radius of curvature of magnetic field-lines, 
$\eta_i = d\ln T_i/d\ln n_e$, and $\eta_e=d\ln T_e/d\ln n_e$. All quantities are evaluated at the rational surface.
The first term on the right-hand side of the previous equation governs the intrinsic stability of the island chain, the
second term parameterizes the effect of the perturbed bootstrap current on island stability, the third
term parameterizes the effect of magnetic field-line curvature on island stability, and the final term parameterizes the
effect of the perturbed ion polarization current on island stability. It can be seen that the perturbed bootstrap current is destabilizing when 
$\eta_i<5.81\,[1+(1+0.283\,\eta_e)\,T_e/T_i]$, magnetic field-line curvature is always stabilizing, and the perturbed ion polarization current
is stabilizing  provided that $0<\eta_i<5.81$.

In the weak neoclassical ion poloidal flow damping regime,  the island width evolution equation of a locked island
chain takes the form (\ref{ew}) 
when the neoclassical ion poloidal flow damping
rate greatly exceeds the neoclassical ion perpendicular flow damping rate.  However, in the opposite limit, 
the island width evolution equation becomes 
\begin{align}\label{ew1}
0.823\,\tau_R\,\frac{d}{dt}\!\left(\frac{W}{r_s}\right)& = {\mit\Delta}'\,r_s \nonumber\\[0.5ex]
&\phantom{=}+15.41\,\epsilon_s^{\,1/2}\,\beta_p\left(\frac{L_q}{L_n}\right)\left(\frac{r_s}{W}\right)
\left[(1-0.172\,\eta_i)+(1+0.283\,\eta_e)\,\frac{T_{e}}{T_{i}}\right]\nonumber\\[0.5ex]
&\phantom{=}-12.64\,\beta_p\left(\frac{L_q^{\,2}}{L_n\,L_c}\right)\left(\frac{r_s}{W}\right)\left[(1+\eta_i)+(1+\eta_i)\,\frac{T_e}{T_i}\right]\nonumber\\[0.5ex]
&\phantom{=} -209.3\,\beta_p\left(\frac{L_q}{L_n}\right)^{\,2}\left(\frac{\rho_i}{W}\right)^{\,2}\left(\frac{r_s}{W}\right)P\left(\frac{\rho_i}{W}\right)\eta_i\,(1-1.367\,\eta_i).
\end{align}
It can be seen that, in this case,  the ion polarization term is modified
in such a manner that it is stabilizing when $0<\eta_i<0.73$. 
This result gives rise to the interesting possibility that,  when $0.73<\eta_i<5.81$, the polarization current can have a stabilizing effect on a
freely rotating island chain, but a destabilizing effect on a corresponding locked chain. This may help to explain the common experimental observation that locked magnetic island chains grow to anomalously 
large widths compared to similar freely rotating chains. 

In the strong neoclassical ion poloidal flow damping regime,  the island width evolution equation of a freely rotating island
chain takes the form 
\begin{align}\label{estrong1}
0.823\,\tau_R\,\frac{d}{dt}\!\left(\frac{W}{r_s}\right)& = {\mit\Delta}'\,r_s \nonumber\\[0.5ex]
&\phantom{=}+15.41\,\epsilon_s^{\,1/2}\,\beta_p\left(\frac{L_q}{L_n}\right)\left(\frac{r_s}{W}\right)
\left[(1-0.172\,\eta_i)+(1+0.283\,\eta_e)\,\frac{T_{e}}{T_{i}}\right]\nonumber\\[0.5ex]
&\phantom{=}-12.64\,\beta_p\left(\frac{L_q^{\,2}}{L_n\,L_c}\right)\left(\frac{r_s}{W}\right)\left[(1+\eta_i)+(1+\eta_i)\,\frac{T_e}{T_i}\right]\nonumber\\[0.5ex]
&\phantom{=} -88.32\,{\cal E}\,\beta_p\left(\frac{L_q}{L_n}\right)^{\,2}\left(\frac{\rho_i}{W}\right)^{\,2}\left(\frac{r_s}{W}\right)P\left(\frac{\rho_{\theta\,i}}{W}\right)(1-0.172\,\eta_i)^{\,2}\,v_0\,(1-v_0),
\end{align}
where
\begin{align}
{\cal E}& = \left(\frac{q_s}{\epsilon_s}\right)^{\,2},\\[0.5ex]
v_0 &= \left(\frac{1+\tau_0}{2}\right)\left(1+\frac{D}{\mu}-\left[1-2\,\frac{D}{\mu}\left(\frac{1-\tau_0}{1+\tau_0}\right)+\left(\frac{D}{\mu}\right)^{\,2}\right]^{1/2}\right),\label{v01}
\\[0.5ex]
\tau_0&=\left(\frac{1+\eta_e}{1-0.172\,\eta_i}\right)\frac{T_e}{T_i}.\label{ev02}
\end{align}
Here, $D/\mu$ is the ratio of the perpendicular particle diffusivity to the perpendicular ion momentum diffusivity at the rational surface, $q_s$ the safety-factor at the
rational surface, and
$\rho_{\theta\,i}=(q_s/\epsilon_s)\,\rho_i$  the poloidal ion gyroradius. 
Note that $0\leq v_0\leq 1$. It can be seen, by comparison with Eq.~(\ref{ew}), that in the strong neoclassical ion poloidal flow damping regime the ion polarization term is enhanced by a factor ${\cal E}=(q_s/\epsilon_s)^{\,2}$ compared to that in the weak neoclassical ion poloidal flow damping regime.\cite{rob,re} Moreover, the polarization term is
always stabilizing (except if $\eta_i= 5.81$, when it is zero). [Note, incidentally, that the expression for the ion polarization term appearing 
in Eq.~(\ref{estrong1}) only holds in the strong neoclassical ion perpendicular flow damping regime, discussed in Sect.~\ref{sss}. 
In the weak neoclassical ion perpendicular flow damping regime, discussed in Sect.~\ref{www}, the expression for the
polarization term is similar in nature, but much  more complicated in form.]

Finally, in the strong neoclassical ion poloidal flow damping regime,  the island width evolution equation of a locked island
chain takes the form
\begin{align}
0.823\,\tau_R\,\frac{d}{dt}\!\left(\frac{W}{r_s}\right)& = {\mit\Delta}'\,r_s \nonumber\\[0.5ex]
&\phantom{=}+15.41\,\epsilon_s^{\,1/2}\,\beta_p\left(\frac{L_q}{L_n}\right)\left(\frac{r_s}{W}\right)
\left[(1-0.172\,\eta_i)+(1+0.283\,\eta_e)\,\frac{T_{e}}{T_{i}}\right]\nonumber\\[0.5ex]
&\phantom{=}-12.64\,\beta_p\left(\frac{L_q^{\,2}}{L_n\,L_c}\right)\left(\frac{r_s}{W}\right)\left[(1+\eta_i)+(1+\eta_i)\,\frac{T_e}{T_i}\right]\nonumber\\[0.5ex]
&\phantom{=} -105.6\,{\cal E}\,\beta_p\left(\frac{L_q}{L_n}\right)^{\,2}\left(\frac{\rho_i}{W}\right)^{\,2}\left(\frac{r_s}{W}\right)\eta_i\,(1-1.367\,\eta_i)\\[0.5ex]
&\phantom{=} -88.32\,{\cal E}\,\beta_p\left(\frac{L_q}{L_n}\right)^{\,2}\left(\frac{\rho_i}{W}\right)^{\,2}\left(\frac{r_s}{W}\right)\left[P\left(\frac{\delta_\perp}{W}\right)-P\left(\frac{\rho_{\theta\,i}}{W}\right)\right]\nonumber\\[0.5ex]
&\phantom{=-}\times(1-0.172\,\eta_i)^{\,2}\,v_0\,(1-v_0),
\end{align}
Here, $\delta_\perp = (q_s/\epsilon_s)\,(2\,\mu_{\perp\,i}/3\,n_e\,m_i\,\nu_{\perp\,i})^{1/2}$, where $\mu_{\perp\,i}$ is the ion perpendicular
viscosity, and $\nu_{\perp\,i}$ the neoclassical ion perpendicular damping rate. 
Roughly speaking, in this case, the ion polarization term is modified
in such a manner that it is only  stabilizing when $0<\eta_i\ltapp 0.75$. (See Fig.~\ref{f3b}).
This result again gives rise to the interesting possibility that,  when $\eta_i\gtapp 0.75$, the polarization current can have a stabilizing effect on a
freely rotating island chain, but a destabilizing effect on a corresponding locked chain. 

According to fluid theory, the enhancement factor of the perturbed ion polarization current in the strong neoclassical ion poloidal flow damping regime
is ${\cal E}=(q_s/\epsilon_s)^{\,2}$.\cite{rob,re} However, it should be noted that, according to kinetic theory,\cite{wil,wil1,wil2} there exists an intermediate neoclassical ion poloidal
flow damping regime in which the enhancement factor is reduced to $\epsilon_s^{\,3/2}\,(q_s/\epsilon_s)^{\,2}$. In this intermediate regime, only the
contribution of the trapped ions to the polarization current is enhanced. It is possible to crudely incorporate the intermediate flow damping
regime into our analysis by writing the enhancement factor in the form\,\cite{sl}
\begin{equation}\label{emod}
{\cal E} = \epsilon_s^{\,3/2}\left(\frac{|v| + \hat{\nu}_i}{|v| + \epsilon_s^{\,3/2}\,\hat{\nu}_i}\right)\left(\frac{q_s}{\epsilon_s}\right)^{\,2}.
\end{equation}
Here, $\hat{\nu}_i=\nu_i/(\epsilon_s\,k_\theta\,V_{\ast\,i})$, where $\nu_i$ is the ion collision frequency, and $v$ is the island phase-velocity parameter
defined in Eq.~(\ref{e44}). 
For the case of a freely rotating island chain,
\begin{equation}\label{emod1}
v =\left(\frac{1-0.172\,\eta_i}{1+\eta_i}\right)v_0.
\end{equation}
On the other hand, for the case of a locked island chain,
\begin{equation}
v = \frac{1-1.367\,\eta_i}{1+\eta_i}.
\end{equation}

\begin{figure}
\epsfysize=4in
\centerline{\epsffile{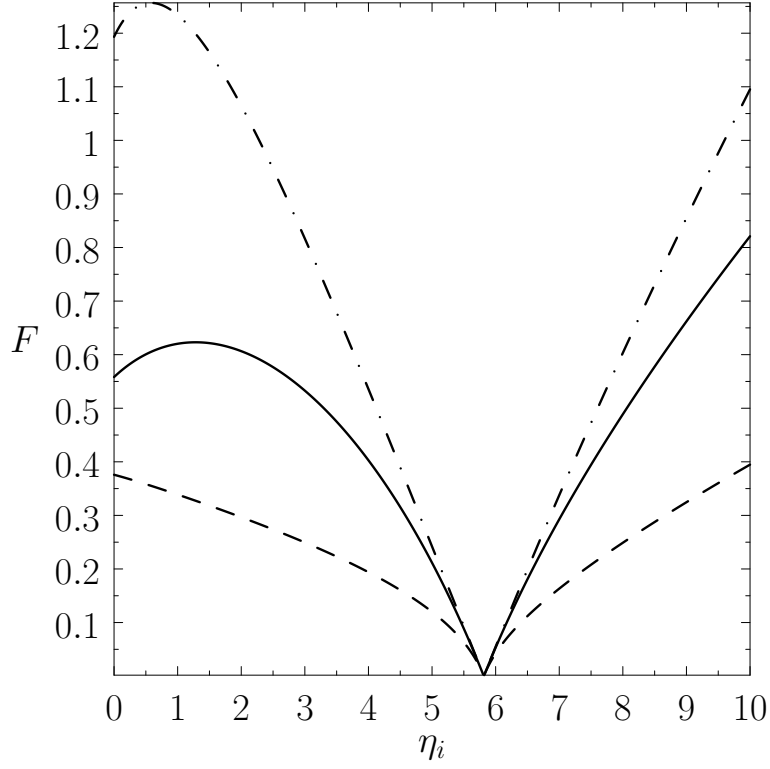}}
\caption{The threshold neoclassical island width function, $F$, calculated as a function of $\eta_i$. The
dashed, solid, and dash-dotted curves correspond to $\hat{\nu}_i=0.1$, $1.0$, and $10.0$, respectively. The
other calculation parameters are $\eta_e=\eta_i$,  $T_e/T_i=1$, $D/\mu=1$,  $L_q/L_c=0$, $\epsilon_s=0.1$ }\label{f4}
\end{figure}

For the case of a freely rotating magnetic island chain in the strong neoclassical ion poloidal flow damping regime (which is the regime that is most relevant to experiments), Eqs.~(\ref{estrong1}), (\ref{emod}), and
(\ref{emod1}) yield the following expression for the threshold island width above which a neoclassical tearing mode grows to large amplitude:
\begin{equation}
W_{\rm crit} = 2.39\,\rho_{b\,i}\left(\frac{L_q}{L_n}\right)^{1/2}P^{\,1/2}\!\left(\frac{\rho_{\theta\,i}}{W_{\rm crit}}\right)F(\eta_e,\eta_i,T_e/T_i,D/\mu, L_q/L_c,\hat{\nu}_i, \epsilon_s),
\end{equation}
where
\begin{align}
F&  = \sqrt{G\,H},\\[0.5ex]
G &= \frac{|1-0.172\,\eta_i|\,v_0 + \hat{\nu}_i\,(1+\eta_i)}{|1-0.172\,\eta_i|\,v_0 + \epsilon_s^{\,3/2}\,\hat{\nu}_i\,(1+\eta_i)},\\[0.5ex]
H&=\frac{(1-0.172\,\eta_i)^{\,2}\,v_0\,(1-v_0)}{(1-0.172\,\eta_i)+(1+0.283\,\eta_e)\,(T_e/T_i) -0.820\,(L_q/\epsilon_s^{\,1/2}L_c)\,[(1+\eta_i)+
(1+\eta_e)\,(T_e/T_i)]},
\end{align}
and $\rho_{b\,i}=\epsilon_s^{\,1/2}\,(q_s/\epsilon_s)\,\rho_i$ is the ion banana width. 
Here, $v_0$ is specified in Eqs.~(\ref{v01}) and (\ref{ev02}), whereas $P$ is specified in Eq.~(\ref{pdef}).  Figure~\ref{f4} shows $F$ calculated as a function of $\eta_i$ for
various different values of $\hat{\nu}_i$. In all cases, it can be seen that $F$ falls to zero at $\eta_i=5.81$. As is clear from Eq.~(\ref{emod1}), this is the
critical value of $\eta_i$ above which the island rotation, relative to the local guiding center fluid, switches from the ion to the electron
diamagnetic direction. Such a switch is likely to trigger a neoclassical tearing mode (because the threshold island width falls to
zero as the switch occurs). A reduction in collisionality (i.e., in $\hat{\nu}_i$) is also likely to trigger a neoclassical
tearing mode (because the threshold island width is a decreasing function of $\hat{\nu}_i$). 

\section*{Acknowledgements}
This research was funded by the U.S.\ Department of Energy under contract DE-FG02-04ER-54742.

\appendix
\section{Useful Integrals}
Let $k=[(1+{\mit\Omega})/2]^{\,1/2}$. Then
\begin{equation}
{\cal A}(k) \equiv 2\,k\,\langle 1\rangle =\left\{
 \begin{array}{lll}
(2/\pi)\,k\,K(k)&\mbox{\hspace{0.5cm}}&0\leq k\leq 1\\[0.5ex]
(2/\pi)\,K(1/k)&&k>1\end{array}
\right.,
\end{equation}
and
\begin{equation}
{\cal B}(k) \equiv \langle |X|\rangle =\left\{
 \begin{array}{lll}
(2/\pi)\,\sin^{-1}(k)&\mbox{\hspace{0.5cm}}&0\leq k\leq 1\\[0.5ex]
1&&k>1\end{array}
\right.,
\end{equation}
and
\begin{equation}
{\cal C}(k) \equiv \frac{\langle X^{\,2}\rangle}{2\,k} =\left\{
 \begin{array}{lll}
(2/\pi)\,[E(k)+(k^{\,2}-1)\,K(k)]/k&\mbox{\hspace{0.5cm}}&0\leq k\leq 1\\[0.5ex]
(2/\pi)\,E(1/k)&&k>1\end{array}
\right.,
\end{equation}
and
\begin{equation}
{\cal D}(k) \equiv\frac{\langle |X|^{\,3}\rangle}{4\,k^{\,2}}   =\left\{
 \begin{array}{lll}
(2/\pi)\,\sin^{-1}(k)\,[1-1/(2\,k^{\,2})]&\mbox{\hspace{0.5cm}}&0\leq k\leq 1\\[0.5ex]
1-1/(2\,k^{\,2})&&k>1\end{array}
\right.,
\end{equation}
and
\begin{equation}
{\cal E}(k) \equiv \frac{\langle X^{\,4}\rangle}{8\,k^{\,3}} =\left\{
 \begin{array}{lll}
(2/3\pi)\left[2\,(2-1/k^{\,2})\,E(k) +(3\,k^{\,2}-5+2/k^{\,2})\, K(k)\right]/k&\mbox{\hspace{0.5cm}}&0\leq k\leq 1\\[0.5ex]
(2/3\pi)\left[2\,(2-1/k^{\,2})\,E(1/k) -(1-1/k^{\,2})\, K(1/k)\right]&&k>1
\end{array}\right..
\end{equation}
Here,
\begin{align}
E(k) &= \int_0^{\pi/2}(1-k^{\,2}\,\sin^2 u)^{1/2}\,du,\\[0.5ex]
K(k) &= \int_0^{\pi/2}(1-k^{\,2}\,\sin^2 u)^{-1/2}\,du
\end{align}
are standard  complete elliptic integrals.\cite{ab}

The following integrals are useful:
\begin{align}
I_1 &\equiv \int_0^\infty\frac{4\,[(2\,k^{\,2}-1)\,{\cal A}-2\,k^{\,2}\,{\cal C}]^{\,2}}{{\cal A}}\,dk = 0.823,\\[0.5ex]
I_2 &\equiv \frac{2\pi}{3}-\int_1^\infty \frac{4}{{\cal C}}\left(\frac{{\cal E}\,{\cal A}}{{\cal C}^{\,2}}-1\right)dk=1.38,\\[0.5ex]
I_3 &\equiv \int_1^{\infty} 16\left(\frac{{\cal D}}{{\cal C}}-\frac{1}{{\cal A}}\right)k^{\,2}\,dk = 1.58,\\[0.5ex]
I_4 &\equiv \int_1^\infty \frac{8}{{\cal C}}\left(1-\frac{1}{{\cal A}\,{\cal C}}\right)dk = 0.357,\\[0.5ex]
I_5&\equiv \frac{4}{2^{\,1/4}}\int_0^\infty \frac{dx}{1+x^{\,4}}=2^{\,1/4}\,\pi=3.74.
\end{align}

\newpage

\end{document}